\documentclass[10pt]{article}

\usepackage{amsmath}
\usepackage{amssymb}
\usepackage{graphicx} 
\usepackage{cite}
\usepackage{color} 

\usepackage{hyperref}

\usepackage[labelfont=bf,labelsep=period,justification=raggedright]{caption}

\makeatletter
\renewcommand{\@biblabel}[1]{\quad#1.}
\makeatother

\date{}

\begin{document}
\begin{flushleft}
{\Large 
\textbf{Modeling competing endogenous RNAs networks}
}
\\
Carla Bosia$^{1\ast}$, Andrea Pagnani$^1$, Riccardo Zecchina$^{1,2}$
\\
{\bf 1} The Human Genetics Foundation (HuGeF), Via Nizza
52, I-10126, Torino, Italy.
\\
{\bf 2} Physics Department and Center for Computational Sciences, Politecnico Torino, Corso Duca degli Abruzzi 24, 10129, Torino, Italy.
\\
$\ast$ E-mail: carla.bosia@hugef-torino.org 
\end{flushleft}



\section*{Abstract}
MicroRNAs (miRNAs) are small RNA molecules, about 22 nucleotide long,
which post-transcriptionally regulate their target messenger RNAs
(mRNAs). They accomplish key roles in gene regulatory networks,
ranging from signaling pathways to tissue morphogenesis, and their
aberrant behavior is often associated with the development of various
diseases.  Recently it has been shown that, in analogy with the better
understood case of small RNAs in bacteria, the way miRNAs interact
with their targets can be described in terms of a titration mechanism
characterized by threshold effects, hypersensitivity of the system
near the threshold, and prioritized cross-talk among targets. The
latter characteristic has been lately identified as competing
endogenous RNA (ceRNA) effect to mark those indirect interactions
among targets of a common pool of miRNAs they are in competition for.
Here we analyze the equilibrium and out-of-equilibrium properties of a
general stochastic model of $M$ miRNAs interacting with $N$ mRNA
targets. In particular we are able to describe in details the peculiar
equilibrium and non-equilibrium phenomena that the system displays
around the threshold: (i) maximal cross-talk and correlation between
targets, (ii) robustness of ceRNA effect with respect to the model's
parameters and in particular to the catalyticity of the miRNA-mRNA
interaction, and (iii) anomalous response-time to external
perturbations.




\section*{Introduction}

A recently discovered molecular mechanism \cite{Arvey10}, lately named
Competing Endogenous RNA (ceRNA) effect \cite{Salmena11,Tay11}, points
out the importance of indirect interactions among transcript RNAs in
competition for the same pool of microRNAs (miRNAs). MiRNAs are small
-- about 22 nucleotide long -- non-coding RNAs which
post-transcriptionally interact with their targets in a sequence
dependent manner. In their mature stage, miRNAs get included in a
RNA-induced silencing complex (RISC) and, eventually, thanks to a 6-8
nucleotide long seed region, bind specifically the miRNA response
elements (MREs) in the 3'UTR of their target mRNAs. Depending on the
degree of complementarity of the seed region, miRNAs can either cleave
the transcripts (large overlap with the target) or downregulate their
translation (low overlap with the target): in either case the net
effect is a reduced amount of mRNAs or proteins. MiRNAs are known to
regulate a multitude of different processes ranging from
differentiation to neural plasticity, and their misfunctioning is
often associated with the development of diseases
\cite{winter2009many, krol_Nature_2010}.

In a nutshell the idea behind the ceRNA effect boils down to the
simple observation that, while interacting with a target mRNA, a
single miRNA cannot act on other targets.  Mature miRNAs ({\em i.e}
miRNAs loaded in RISC) are thus the limiting factor in a system of
potentially interacting target mRNAs. If for example gene A which
shares one miRNA with gene B, is up-regulated the common miRNAs will
tend to bind preferentially to mRNA A due to its increased
concentration. Consequently, mRNA of gene B will be less repressed
resulting in a subsequent increased concentration
\cite{seitz2009redefining, Arvey10, Salmena11,Tay11,
  karreth2011vivo}. Other studies have independently provided further
evidences for miRNA mediated trans-regulatory mRNA effects
\cite{Jeyapalan2011, sumazin2011extensive}.  Since each miRNA can have
several targets, a complex indirect interaction network among
different targets emerges, where nodes are mRNA transcripts and there
is a link between two nodes if they have at least one miRNA in
common. Then, the highest the number of common miRNAs or MREs, the
strongest the link. Such crosstalk effect has been observed in
bacteria where the role of miRNAs is played by small RNAs (sRNAs) and
it is due to a titrative interaction among sRNAs and targets
\cite{Levine07}.  Depending on the number of sRNA binding elements
crosstalk among sRNA targets can then be prioritized and selective
\cite{Levine07,Mitarai07}.

Interaction via titration mechanisms entails a threshold-like behavior
between the two interacting molecules, where the threshold position is
determined by the relative amount of them
\cite{Elf03,Buchler08,Levine07,Levine08, Shimoni07}.  This means that
as long as the concentration of one of these two molecules is below
the threshold almost all of them are bound in complexes with the
second ones and their free amount is very low. Increasing their
concentration beyond the threshold results in an increased amount of
free molecules, while the others will be in turn almost all bound in
complexes. Moreover, systems of molecules interacting in a titrative
fashion also show a hypersensitivity in proximity to the threshold to
changes in the molecule production rates \cite{Elf03,Buchler08}.  In
particular controlled conditions it has been shown that it is right
near the threshold, where sensitivity is maximal, that crosstalk among
sRNA targets is maximal too \cite{Levine07}.

Remarkably, Mukherji and co-workers \cite{Mukherji11} recently
observed a threshold-like effect also in miRNA target expression in
single cells.  Moreover, in line with studies in bacteria
\cite{Levine07,Levine08} and with earlier works on protein-protein
interaction \cite{Elf03, Buchler08}, they tested a mathematical
deterministic model of molecular titration to describe their results
and found it in good agreement with experimental observations. Such
results strengthen the idea that behind the ceRNA effect there is a
miRNA-target titration mechanism.

Motivated by \cite{Mukherji11} and \cite{Salmena11, Tay11} and by
results obtained in experiments with bacteria
\cite{Levine07,Mitarai07,Levine08}, in this paper we extend previous
models to the case of a general network of $M$ miRNAs titratively
interacting with $N$ target mRNAs (ceRNAs) and analyze it from a
stochastic point of view.  So far analytical predictions from models
for titrative interactions did not go beyond the mean-field limit
\cite{Levine07,Levine08,Mukherji11} or were limited to the case of
small circuits because of the nonlinearities involved \cite{Elf03}.
However, (i) stochasticity plays a central role in gene expression
mostly when numbers of molecules involved are modest \cite{Kaern05,
  Maheshri07, Raj08} and (ii) small circuits are usually embedded in
more complex networks so that induced interactions might be
relevant. Since potential crosstalk among miRNA targets is effective
right in proximity to the threshold, where free chemical species ({\em
  i.e.} not bound in complexes) are present in small numbers, it is
necessary a stochastic analysis of the system.

Here we show that, despite the complexity and the intrinsic
non-linearity of the system,  a shrewd use of the moment generating
function approach plus a simple Gaussian approximation are enough to
obtain analytical expressions for noise and Pearson's correlation
coefficients for all the molecular species considered in a generic
network.

As a preliminary result we describe, at the level of the independent
molecular species approximation ({\em viz.} mean-field), the onset of
a threshold-like behavior typical of titration mechanism
\cite{Elf03,Buchler08,Levine07,Levine08, Shimoni07}, which has been
specifically investigated in \cite{pandolfi12} in the case of a
miRNA-mediated mRNA interaction, and discuss the possible mechanism
leading to a specificity of the interactions.

Secondly, for the first time, we derive analytical results beyond the
independent molecular species approximation which allows for the
characterization of profiles for means, noise and Pearson's
correlation coefficients, comparing them with numerical simulations.
Interestingly, we found that in proximity to the threshold both noise
and correlation profiles among the different molecular species (in
terms of Fano factor and coefficient of variation) show a maximum.
Titration-like interactions could thus be an adequate mechanism to
maintain homeostasis in the system: even if the noise increases,
ceRNAs or miRNAs fluctuate in a highly correlated manner as discussed
in \cite{Li09, Inui10, Ebert12}.

Among the different parameters characterizing miRNA-mRNA interactions,
the degree of catalyticity -- {\em i.e.} the fraction of mRNA
molecules that are recycled after the interaction with their target --
is among the most disputed yet less understood ones:
\cite{Hutvagner02,Haley04} support an almost completely catalytic
interaction ($\alpha \sim 0$), while at the opposite range
\cite{Liu05, Pillai05, Kai10} support an almost completely
stoichiometric interaction ($\alpha \simeq 1$). Finally, intermediate
values of catalyticity are indeed supported by a recent work
\cite{Baccarini11}. Here we show that ceRNA effect is robust with
respect to this parameter too.  In the limiting case of a completely
catalytic interaction ({\em i.e.}  $100\%$ of the miRNA is recycled) a
threshold-behavior is still observed as an intrinsically
out-of-equilibrium phenomenon: the location of the threshold turns out
to be a monotonously increasing function of time such that, at
equilibrium (long-time limit), no threshold behavior is observed.

An out-of-equilibrium characteristic of the system predicted by the
model is the response time of a ceRNA embedded in a network after a
single factor perturbation. Again, in proximity to the threshold, we
observe peculiar trends: upon switching on or off another ceRNA in the
network the response times show a maximum and a minimum respectively,
and the qualitative profiles are independent of the number of ceRNAs
in competition.

Finally we conclude proposing a series of specific experiments aiming
at validating both qualitatively and quantitatively the model's
predictions.


\section*{Results}


\subsection*{Definition of a network of interaction miRNAs-ceRNAs}

The network we are interested in describing is schematically depicted
in Figure \ref{schema}A, where $M$ different free mature miRNAs
(colored stars) can interact with $N$ different free target mRNAs
(colored pentagons). miRNAs and target mRNAs interact via a
titration-like mechanism \cite{Mukherji11}. As a first approximation
we can think the mRNAs as irreversibly lost due to the miRNAs actions
(miRNA-target association rate much greater than dissociation rate)
while the miRNAs can eventually be recycled. Figure \ref{schema}B
shows a cartoon of such mechanism in which two different DNA molecules
(green rectangles) are transcribed with rates $k_{S_i}$ and $k_{R_j}$
to become miRNA $S_i$ and mRNA $R_j$ respectively. Eventually $S_i$
and $R_j$ either degrade (broken gray stars and pentagons) with rates
$g_{S_i}$ and $g_{R_j}$ or interact binding in a complex $C_{ij}$ via
an effective association rate $g_{ij}$.

The effective association rate $g_{ij}$ should be thought as a
combination of association, dissociation and degradation rates of the
miRNA-mRNA complex $C_{ij}$ (see SI for more details).  Once in the
complex the mRNA $R_j$ cannot be translated or utilized anymore. The
parameter $\alpha$ (with $0 \leq \alpha \leq 1$) is a measure of the
catalyticity of the miRNA, that is the ability the miRNA has to be
available again once having interact with its target.  Thus, $\alpha =
1$ means that for each mRNA $R_j$ bound in a complex $C_{ij}$ there is
also one miRNA $S_j$ sequestered (and no more able to interact with
its other targets) while $\alpha = 0$ implies that mRNA $R_j$
effective degradation is increased by $g_{ij}$ but this does not have
any effect on the miRNA $S_i$.
 
\subsection*{Mean field approximation: threshold behavior and cross-talk}

The onset of a threshold-like response as a consequence of a titration
mechanism is a rather well known phenomenon
\cite{Elf03,Buchler08,Levine07, Levine08, Shimoni07,
  Mukherji11,pandolfi12}. In Figures \ref{duale_cerna}A and
\ref{duale_mirna}A, we show an example of threshold effect in the case
$M=N=2$ as a function of different ceRNA and miRNA
concentrations. Such an effect can be derived under the assumption
that the joint probability distributions of the different molecular
species are statistically independent, as explained in Section {\em
  Materials and Methods}.

In a general network of interaction of $N$ ceRNAs and $M$ miRNAs, when
miRNA-target interaction strength is high, following the derivation of
Eq.~\ref{soglia} and depending on the control parameter we decide to
tune, two distinct phases emerge: (i) if all target transcription
rates are below the threshold level, explicitly computable in terms of
all other model's parameters, all targets turn out to be bounded in
complexes and the free molecule ({\em~i.e.} not bounded) share is very
low, (ii) if at least one of the transcription rate -- say the q-th
target -- is above threshold, then all other target free molecule
shares are expressed in finite amount. As shown in Figure
\ref{duale_cerna}A, the emerging scenario entails a cross-talk
mechanism where a single mRNA target above threshold is able to drive
the other common mRNA targets above threshold. The hypothesis of a
strong ceRNA-miRNA interaction can be relaxed, and still, a smoother
threshold-like behavior is observed \cite{Levine07}.

Interestingly enough we note that if, as control parameter, we decide
to tune the p-th miRNA transcription rate, keeping all the remaining
model's parameters fixed, a mirror-like scenario emerges (as displayed
in Figure \ref{duale_mirna}A): in complete analogy with the case
previously discussed, also miRNAs cross-talk through ceRNAs.  Here
again, as long as all miRNAs transcription rates are below threshold,
free miRNA molecule shares are very low.  As the first miRNA
transcription rate crosses the threshold, all other miRNAs show a
substantial increase of their free share. In this case too there is a
clear cross-talk between miRNAs. It is interesting to note that the
threshold value predicted by the model (see section {\em Materials and
  methods}) occurs at near-equimolar concentrations of the different
chemical species.

If a hierarchy is present for the miRNA-target interaction strengths
$g_{ij}/(g_{R_i}g_{S_j})$ \cite{Levine07, pandolfi12}, for
example accounting for different miRNA regulatory elements (MREs) for
different target mRNAs, then a hierarchy will be also established in
the other target (miRNA) signal amplification levels when the amount
of target mRNAs (miRNAs) is moved from below to above the threshold
value. Targets sharing similar MREs will be more co-regulated than
targets sharing only few MREs \cite{pandolfi12}.  The miRNA-target
interplay may thus be selective depending on the particular affinities
and binding strengths \cite{Levine07,Mitarai07}.  This leads to a
complex regulatory network with non-trivial indirect interactions
among targets in competition for the same pool of miRNAs.

The network sketched in Figure \ref{schema}A is a crude simplification
of what should be a real-case ceRNA's network. To make things slightly
more realistic see Figure \ref{select}A, where two groups of ceRNAs
interact through two distinct sets of miRNAs
\cite{pandolfi12}. However, a small subset of miRNAs makes the two
groups of ceRNAs, otherwise statistically independent, {\em weakly
  interacting} by cross-connecting the two sets. We simulated the
network's dynamics using the Gillespie algorithm in two different
settings: in the first one, we modulate over time the transcription
rate of one ceRNA, starting with a value below threshold and we first
increase the transcription of one specific ceRNA (ceRNA1) rate after
35 hours. A first observation is that it is enough to bring above
threshold a single ceRNA, to set the whole network in its
non-repressed state. The second observation is that ceRNA-mediated
regulation can be specific, {\em i.e.}  we observe a clear hierarchy
in the response of the different ceRNAs (see Figure \ref{select}B):
those ceRNAs sharing the largest set of miRNA (red pentagons) respond
more then the yellow pentagon set that shares a fewer number of
ceRNAs. A second increase in the transcription rate of ceRNA1 after 70
hours makes the hierarchy in the responses even more
clear. Interestingly, also the sets of ceRNAs (orange and blue
pentagons) which do not share any targeting miRNA respond to the
over-expression of ceRNA1 (although less than the previous two
groups), thanks to an undirected effective interaction: ceRNA1 pulls
up the red and yellow pentagon sets, the yellow pentagon set pulls up
the orange, and the latter the blue pentagon set.

In the second setting (see Figure \ref{select}C), we analyze the
mirror scenario in which miRNA10 transcription rate is
increased. Again the hierarchical responses of the different miRNAs is
clearly visible.

\subsection*{Beyond mean field approximation: noise and correlation coefficients} 

To get insight into molecular species correlations for the miRNA-ceRNA
interaction network we then assume that the joint probability
distribution $P$ for the different molecular species is a multivariate
Gaussian (see section Materials and Methods). This {\em ansatz} turns
out to be useful since all moments of a multivariate Gaussian can be
expressed as a function of the first two, {\em i.e.} in terms of means
and covariances. We will assume that the vector $\vec X =
(X_1,\dots,X_{N+M}) := (R_1,\dots,R_N,S_1,\dots,S_M)$ is distributed
according a Gaussian multivariate measure of mean $\mu_i:= E(X_i)$ and
covariances $c_{ij} := E(X_i X_j) - E(X_i) E(X_j)$. Thus the generic
third and fourth moments read $E(X_i X_j X_k) := c_{ij}\mu_k +
c_{ik}\mu_j + c_{jk}\mu_i$ and $E(X_i X_j X_k X_l) := c_{ij}c_{kl} +
c_{ik}c_{jl} + c_{il}c_{jk}$.

In this way we are able to obtain a closed system of equations for
$\langle X_i \rangle$, $\langle X_{i}^2 \rangle$ and $\langle X_i
X_j\rangle$ (see Supplementary Material for a detailed analysis).
This assumption is not arbitrary (the usual van Kampen's expansion
method \cite{van2007stochastic} shows the master equation is Gaussian
except for small corrections) and interestingly performs better than
the most widely used linear noise approximation (see Supplementary
Materials) when compared with Gillespie's simulations (see
\cite{TESI_SPAGNOLO} for a nice introduction to the subject). 
Under this approximation we then find an analytical expression for means,
noise and Pearson's correlation coefficients.

The threshold is characterized not only by the abrupt change of the
mean quantities as a function of the control parameter, but also by
Pearson's correlation coefficients and noise (both related to the
covariances) which turn out to show a maximum around the
threshold. For each molecular species we evaluated in terms of
variance $\sigma_{\langle x_i\rangle} := \sqrt{\langle X_i^2 \rangle -
  \langle X_i\rangle^2} $ the Fano factor, $f_{x_i} = \sigma_{\langle
  x_i\rangle}^{2}/\langle x_i \rangle$, and the coefficient of
variation, $CV_{x_i} = \sigma_{\langle x_i\rangle}/\langle
x_i\rangle$, which are both measures of noise. While the first one
tells how much a particular process is different from a Poisson
process, the second is a dispersion index.  Figures
\ref{duale_cerna}B,C and \ref{duale_cerna}B,C show such noise profiles
as a function of ceRNA1 or miRNA1 transcription rate. As it is
possible to notice in Figures \ref{duale_cerna}B and
\ref{duale_mirna}B, in proximity to the threshold the joint
probability distributions are far from being independent ($f_{x_i} \gg
1$ for all indexes $i$ labelling the different chemical species) while
a multivariate Gaussian approximation is better suited to describe the
simulation results.  In Figures \ref{duale_cerna}C and
\ref{duale_mirna}C we plot the CV profiles.  Increasing the ceRNA1
(miRNA1) transcription rate we observe a decreasing noise profile for
ceRNAs (miRNAs) and an increasing one for miRNAs (ceRNAs), as expected
because of the increasing and decreasing amount of free ceRNAs (miRNAs)
and miRNAs (ceRNAs) respectively.  Interestingly, right close to the
threshold it is possible to notice a bump in the CV profiles.  This
phenomenon, due to the variances growing faster than means, is
compatible with the bimodal distributions experimentally observed and
verified via simulations in particular controlled conditions in
bacterial sRNA target \cite{Levine, Hao11}.

The Pearson's correlation coefficients, $\rho_{X_i, X_j} =
\frac{\langle x_i x_j\rangle - \langle x_i \rangle \langle x_j
  \rangle}{\sigma_{\langle x_i \rangle } \sigma_{\langle x_j
    \rangle}}$, are shown in Figures \ref{duale_cerna}D and
\ref{duale_mirna}D.  The profile of the curves as a function of the
control parameter, with a well-defined maximum, confirms the system
hypersensitivity near the threshold.  Analogously, we can define the
Pearson correlation coefficient between miRNAs and ceRNAs (not
shown). In this case, miRNAs and ceRNAs are negatively correlated.

It is interesting to notice that exactly where the number of
interacting molecules is small and the noise profiles show local
maxima, the statistical correlation between molecular species is
maximal too. Speculatively, the titration interaction mechanism
provides for a tool able to maintain the network homeostasis:
potentially interacting ceRNAs (or miRNAs) needed in the same time
fluctuate together .


\subsection*{Threshold effect and miRNA-target catalytic interaction}

So far we considered a titrative stoichiometric ($ 0 < \alpha \leq 1$)
ceRNA/miRNA interaction.  However, the open question is if cross-talk
among miRNAs or miRNA targets can be possible in case of purely
catalytic-like interaction (that is, in case of complete miRNA
recycling, or rather $\alpha = 0$ in Equation \ref{master})
\cite{Kai10}.

It is straightforward to see that, at the steady state, equations for
the various $\langle R_j \rangle$ (or $\langle S_i \rangle$) decouple
when $\alpha=0$ (see Equation \ref{indep}) \cite{pandolfi12}. As a
consequence, no cross-talk is possible among ceRNAs (or miRNAs).  We
found that in the out of equilibrium phase instead, the behavior is
different.

We considered the time evolution of the system in Equation 1 of the
Supplementary Material, and then took pictures of the system at a
given time $t$. If $t$ is sufficiently small with respect to the time
the complexes need to reach the steady-state, for different values of
miRNA (or ceRNA) transcription rate we can observe the threshold
behavior of Figure \ref{ooe}A. Consequently ceRNAs or miRNAs
cross-talk is possible, and statistical correlations are maximal, as
shown by the Pearson's correlation coefficient profile in Figure
\ref{ooe}B.

The emerging picture is that of a dynamical threshold
whose value at a given time t tends monotonously to the equilibrium
one in case of $\alpha \neq 0$ and to infinity in case of $\alpha = 0$
for large time. In the latter case no cross-talk is observed at
equilibrium (Figure \ref{ooe}C,D).

The ceRNA effect is therefore robust also in case of catalytic
miRNA-target interaction, the crucial point lieing in the instant of
time at which we look at the system.


\subsection*{Response times}

We have already discussed the threshold effect due to titrative
miRNA-target interaction and how the system displays strong
sensitivity (maximum cross-talk) and the maximal statistical
correlation.  We now want to understand how fast the system responds
to an external perturbation. In particular we want to compute the time
needed for a particular ceRNA (say ceRNA1) to reach the equilibrium
after the instantaneous over-expression or knock-out of a second ceRNA
(ceRNA2).

Following \cite{bosia12}, we consider two different settings: (i) to
mimic a sudden signal which saturates ceRNA2 promoter at $t = 0$, the
transcription rate $k_{R_2}$ of ceRNA2 switches from zero to a given
value ($\mbox{ceRNA2}_{\mathrm{OFF} \rightarrow \mathrm{ON}}$), (ii) to mimic the
opposite condition of a sudden drop of the activating signal at $t =
0$, the transcription rate of ceRNA2 $k_{R_2}$ switches from its
initial value to zero ($\mbox{ceRNA2}_{\mathrm{ON} \rightarrow \mathrm{OFF}}$).

Defining the response time as the time needed to reach half of the way
between initial and final ceRNA1 steady state, we evaluate the
response times for both switch-on ($\mbox{T}_{\mathrm{ON}}$) and
switch-off ($\mbox{T}_{\mathrm{OFF}}$) ({\em i.e.} for
$\mbox{ceRNA2}_{\mathrm{OFF} \rightarrow \mathrm{ON}}$ and
$\mbox{ceRNA2}_{\mathrm{ON} \rightarrow \mathrm{OFF}}$ respectively)
conditions.  We integrated numerically the deterministic system of
equations obtained with $M=1$ and $N=2$ (see Equation 2 in
Supplementary Material) to calculate: (i) the time $T_{\mathrm{ON}}$
such that $R_{1}(T_{\mathrm{ON}}) = R_{1_{0}} + (R_{1_{ss}} -
R_{1_0})/2$ (where $R_{1_{0}}$ and $R_{1_{ss}}$ are the initial and
final ceRNA1 steady-state respectively), (ii) the time
$T_{\mathrm{OFF}}$ such that $R_{1}(T_{\mathrm{OFF}}) = R_{1_{0}} -
(R_{1_{0}} - R_{1_{ss}})/2$.  The initial conditions are $R_2(0) = 0$
and $R_1(0)$ and $S(0)$ with their steady state values in absence of
$R_2$ in the former case, and $R_2(0) \neq 0$ and $R_1(0)$ and $S(0)$
with their steady state values in presence of $R_2$ in the latter. We
also considered a slightly more complex network in which more ceRNAs
are present and we compute ceRNA1 response time with $N=5,10,20$.

We then ask two questions: (i) how the response time of ceRNA1 changes
at different values of basal miRNA concentration, and (ii) what
happens when the system is complicated by the addition of other
competing targets.

As displayed in Figure \ref{tempi}A,B, upon increasing miRNA
transcription rate ceRNA1 $T_{\mathrm{ON}}$ and $T_{\mathrm{OFF}}$
show a maximum and a minimum respectively. Both the maximum and the
minimum are located at the threshold, where ceRNA1 initial and final
equilibrium values are near (see Figure \ref{tempi}C). Such response
time trend suggests an {\em out-of-equilibrium phase transition}, for
which the system experiences anomalous dynamical features around
threshold.  Let us point out that around threshold, despite the change
in terms of number of molecules from initial and final steady state is
small, as depicted in Figure \ref{tempi}C, $T_{\mathrm{ON}}$ is
largely increased while $T_{\mathrm{OFF}}$ is decreased. Moreover, the
qualitative shape of the curve is robust with respect to the number of
targets in competition for the same miRNA (see Figure \ref{tempi}A,B
where different line colors correspond to a different number of ceRNAs
in the interaction's network): the maximum (resp. the minimum) of the
response time depends only mildly on the number of ceRNA competitors,
whereas the location of the threshold at which the free molecule share
of ceRNA1 starts being repressed depends linearly on the number of
competitors.  Moreover, the statistical correlation between ceRNA1 and
ceRNA2 seems independent from the size of the ceRNA's network: the
maximum level of correlation is almost the same upon increasing the
number of ceRNAs with only a shift to higher miRNA transcription rates
(Figure \ref{tempi}D). Therefore ceRNA1 and ceRNA2 are always very
correlated, notwithstanding the dynamical anomalies in the
response-time around threshold.


\section*{Discussion}

In this paper we analyzed the theoretical framework for the stochastic
description of a general network of $M$ miRNAs interacting with $N$
target mRNAs via a titration mechanism. With a dexterous use of the
moment generating function approach plus simple Gaussian approximation
we showed that it is possible to obtain analytical expressions for
means and covariances for all the interacting molecules present in the
system.

We have first shown how the already well understood threshold effect
implied by titrative interaction
\cite{Elf03,Buchler08,Levine07,Levine08, Shimoni07} entails with
interesting cross-talk phenomena which, so far, have been only
partially investigated from the experimental point of view \cite{
  Arvey10, Salmena11,Tay11, karreth2011vivo, Jeyapalan2011,
  sumazin2011extensive}. In particular the issue of the mirror
scenario -- for which not only ceRNAs cross-talk through competing for
the same set of miRNAs, but, symmetrically the same set of miRNAs too
cross-talk through the common set of ceRNA -- is a straightforward
verification of the {\em ceRNA hypothesis} which, at the best of our
knowledge, has never been attempted so far.  In practice, knowing the
set of miRNAs belonging to a specific ceRNA network, one could
knock-down (resp. over-express) a given miRNA in the network. In this
case, the model predicts that the other miRNAs in the network, driven
by the controlled miRNA knock-down (resp. over-expression), should
increase (resp. decrease) their free molecule share. Such an effect
could be directly measurable as a down-regulation
(resp. up-regulation) of any of the miRNAs targets (either belonging
to the same ceRNA network, or to any other secondary target).

In addition to cross-talk and threshold phenomena, the model predicts
interesting and experimentally measurable trends for the noise and
Pearson's correlation coefficient profiles. In proximity to the
threshold, where all the free molecular species involved in the system
are present in small numbers, both the noise measures we analyzed
(Fano factor and coefficient of variation) show a maximum (for the
latter coefficient the maximum is local). These behaviors are
interpretable in terms of bimodal distributions for each molecular
species involved in the titrative mechanism
\cite{Hao11}. Interestingly the bimodality has been experimentally
measured in a simple sRNA-mediated circuit in Bacteria \cite{Levine},
and could be potentially verified in our ceRNA case.

In proximity to such threshold value, also the Pearson's correlation
coefficients among ceRNAs or miRNAs show a maximum, meaning that the
statistical correlation among molecules deriving from different genes
is high. That is, not only the system is hypersensitive to little
changes in the control parameter, but also fluctuations are highly
correlated. As a matter of fact, the titration mechanism of
interaction establishes a positive coupling among ceRNAs belonging to
different genes (or among miRNAs). While the intensity of such
correlation depends mostly on the combination the basal transcription
rates of each particular gene (so that different genes speak each
other at different intensities, but the level of correlation is
established by the particular parameters), the location of the maximum
is a determined by all the molecular species in
competition. Furthermore, such statistical correlation is robust with
respect to the number of ceRNAs involved in the system (with just a
shift in the location of the threshold when increasing the number of
ceRNAs) and also with respect to the catalyticity parameter
$\alpha$. When $\alpha$ is zero, meaning that all the miRNAs are
recycled, it is still possible to observe the threshold effect and the
maximum in correlations' profiles as an out-of-equilibrium
characteristic of the system. Thus, the {\em ceRNA effect} is always
present, provided that the observation's time is short enough.

To investigate experimentally these features, quantitative
fluorescence microscopy seems, for the time being, the most promising
technique. Previous works not directly related to the {\em ceRNA
  hypothesis} (see \cite{Levine07} for a seminal work in bacteria, and
\cite{Mukherji11} in human cell lines) used two-colors fluorescent
reporter systems. The construct typically consists of a bidirectional
drug-inducible promoter driving the expression of the two fluorescent
proteins. The 3'UTR of the fluorescent proteins can be engineered to
control the binding sites, and so the miRNA-mRNA binding affinity for
the targeting miRNAs of interest. Both in \cite{Levine07} and
\cite{Mukherji11}, the method was used to monitor the threshold effect
in a simple sRNA/miRNA $\rightarrow$ mRNA interaction. At the expenses of
creating more complex constructs, an analogous technique could be
deployed to investigate threshold, cross-talk, and noise/correlation
behavior of simple ceRNA networks. In the most straightforward
implementation one needs two reporter constructs: (i) the first
construct consists of a bidirectional reporter system composed by the
3'UTR of ceRNA1 concatenated to the fluorescent gene (say green), and
on the other side a miRNA binding site free 3'UTR concatenated to a
second fluorescent gene (say yellow) to monitor the transcription
activity, (ii) the second construct consists of a single reporter
composed by the 3'UTR of ceRNA2 concatenated with a third fluorescent
gene (say cherry). In this way one could simultaneously monitor the
activity of both ceRNAs (green, cherry) as a function of the
transcriptional activity of ceRNA1 (yellow) which would validate both
qualitatively (in terms of the profile predicted by the model) and
possibly quantitatively (by allowing a multi-parametric fit of the
model's kinetic constants from the experimental data) the model
predictions as displayed, for instance, in Figure \ref{duale_cerna}.

Finally, the model shows interesting out-of-equilibrium features
around threshold which could be experimentally testable. In particular
the peculiar response time profile as a function of the distance from
the threshold could be directly measured by means of quantitative
time-lapse fluorescence microscopy \cite{MuzzeyVanOudenarden2009} and
flow cytometry to monitor ceRNAs dynamics. To monitor the dynamics of
two ceRNAs, one could conservatively construct a two colors
fluorescent reporter system that allows for simultaneous monitoring of
protein levels (see again\cite{Levine07, Mukherji11}). Of course
larger networks could be potentially monitored using multiple colors.


\section*{Materials and Methods}

\subsection*{Stochastic simulations}
Stochastic simulations have been performed via implementation of
Gillespie's first reaction algorithm~\cite{Gillespie76}.

\subsubsection*{Theoretical framework: stochastic model}

\begin{figure}[h]
\includegraphics[width=1\textwidth]{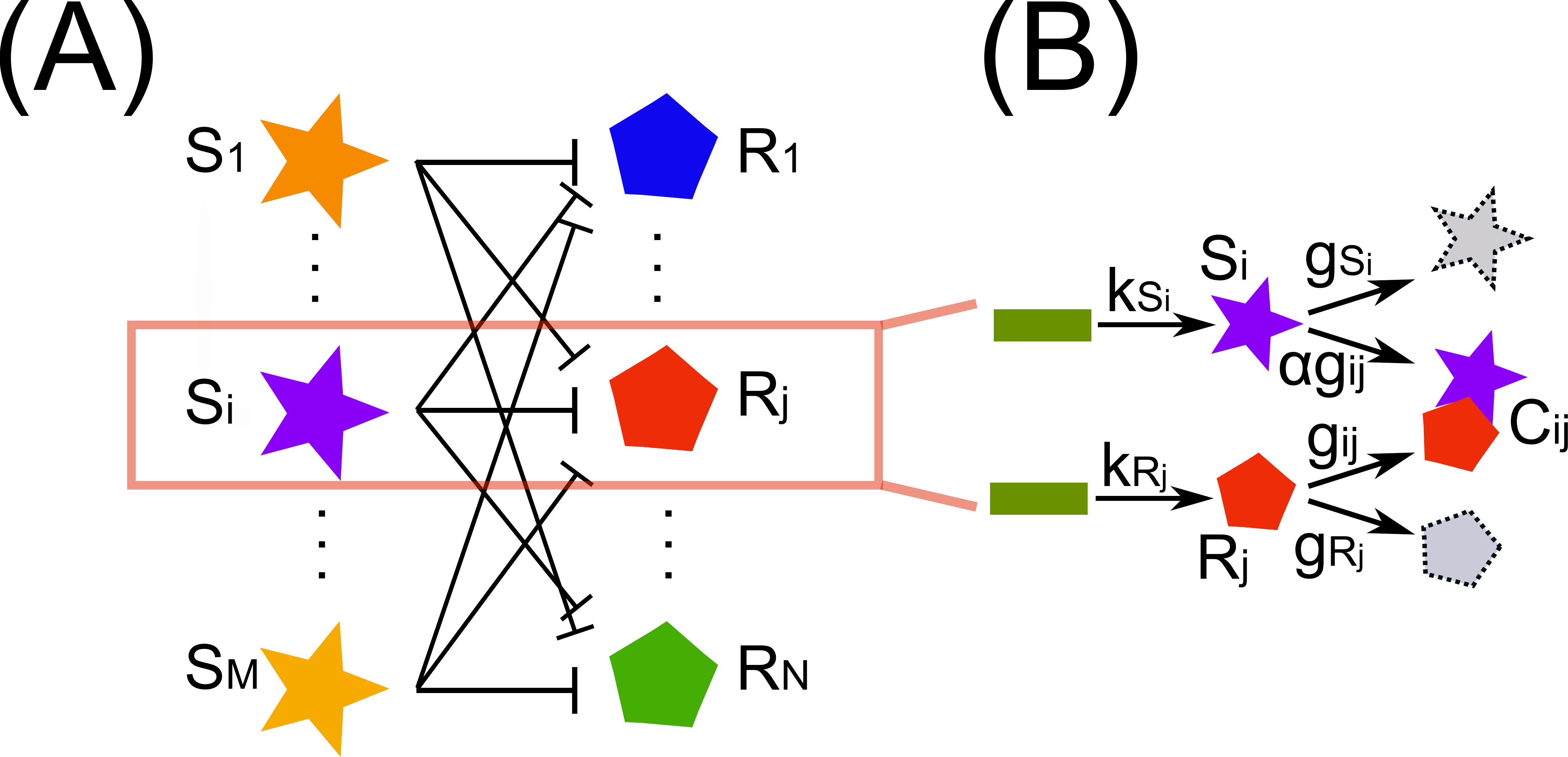}
\caption{{\it Representation of a generic miRNA-target interaction
    network.}  (A) Simplified picture of a miRNA-ceRNA interaction
  network. (B) For each miRNA ($S_i$) and ceRNA ($R_j$) present in the
  network we consider the main steps of transcription (rates $k_{S_i}$
  and $k_{R_j}$ respectively) and degradation (rates $g_{S_i}$ and
  $g_{R_j}$ respectively) plus a titrative interaction between miRNA
  and ceRNA. miRNA and ceRNA can therefore form a complex $C_{ij}$
  with effective association rate $g_{ij}$. The parameter $\alpha$
  (the catalyticity parameter) tells which is the probability a miRNA
  is recycled after having interact with one of its targets.}
\label{schema}
\end{figure}

\begin{figure}[h]
\includegraphics[width=1\textwidth]{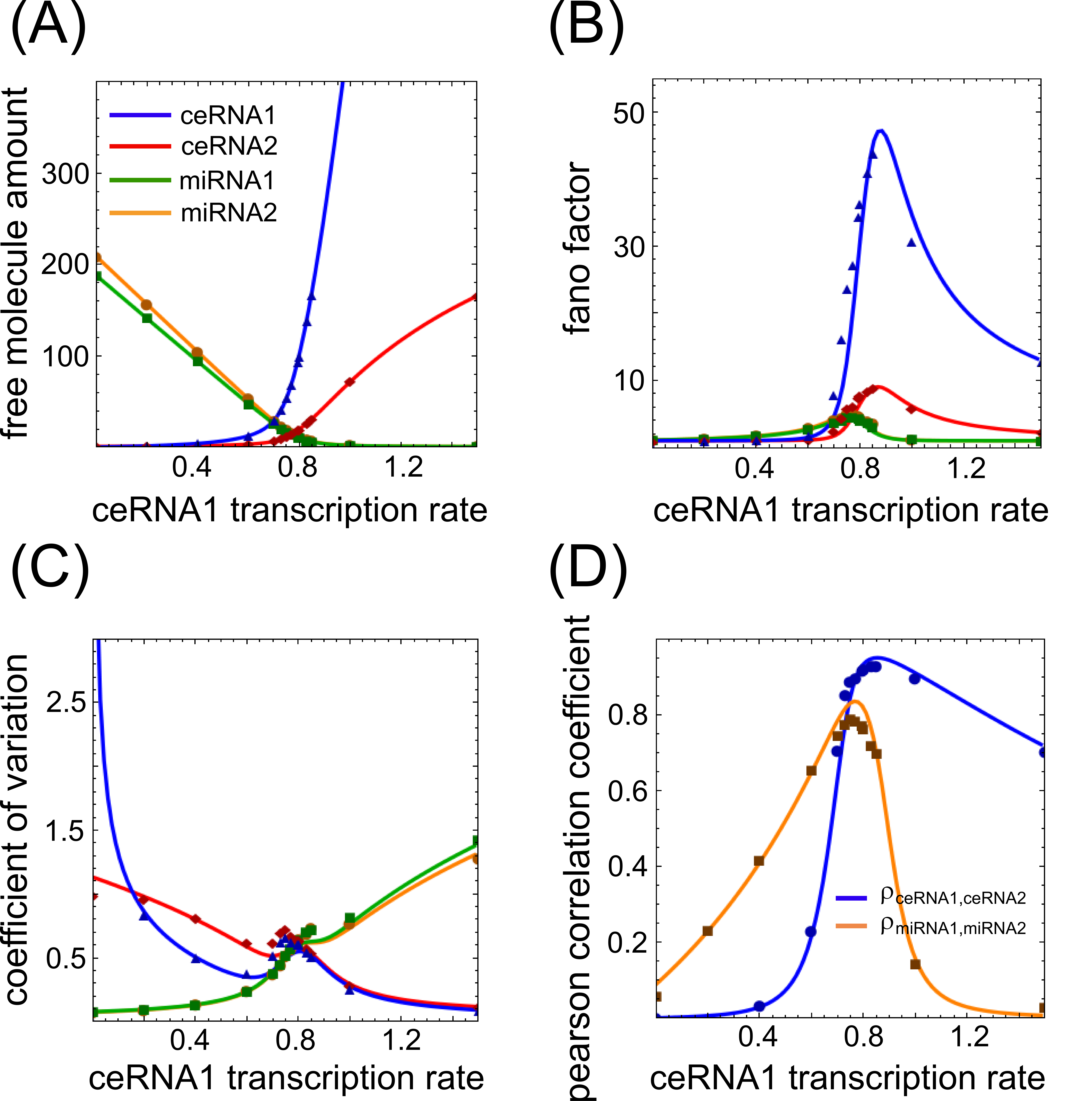}
\caption{{\it Threshold, noise and Pearson's correlation coefficient varying ceRNA transcription rate. }(A-C) Steady state value for means, Fano factors and coefficients of
variation for each free molecular species in a system of two miRNAs
(miRNA1 and miRNA2, green and orange lines respectively) interacting
with two ceRNAs (ceRNA1 and ceRNA2, blue and red lines respectively)
varying the concentration of ceRNA1. In proximity to the threshold the
system shows hypersensitivity to changes in the control parameter
(ceRNA1 transcription rate), captured by a maximum in the Fano factors
(panel B). For the same values of ceRNA1 transcription rate, the local
maximum in the coefficients of variation (panel C) is the fingerprint
of bimodal distributions in the number of molecules for each molecular
species.  (D) Pearson's coefficients between the two miRNAs (orange
line) and the two ceRNAs (blue line). The two lines show a maximum in
proximity to the ceRNA1 transcriptiom rate threshold value, meaning
that there is a region of parameters where the fluctuations in the
number of ceRNAs or miRNAs are highly correlated.  Lines are the
results of Gaussian approximation while symbols are Gillespie's
simulations. For panels B,C the line color-code is the same as in
panel A.}
\label{duale_cerna}
\end{figure}

\begin{figure}[h]
\includegraphics[width=1\textwidth]{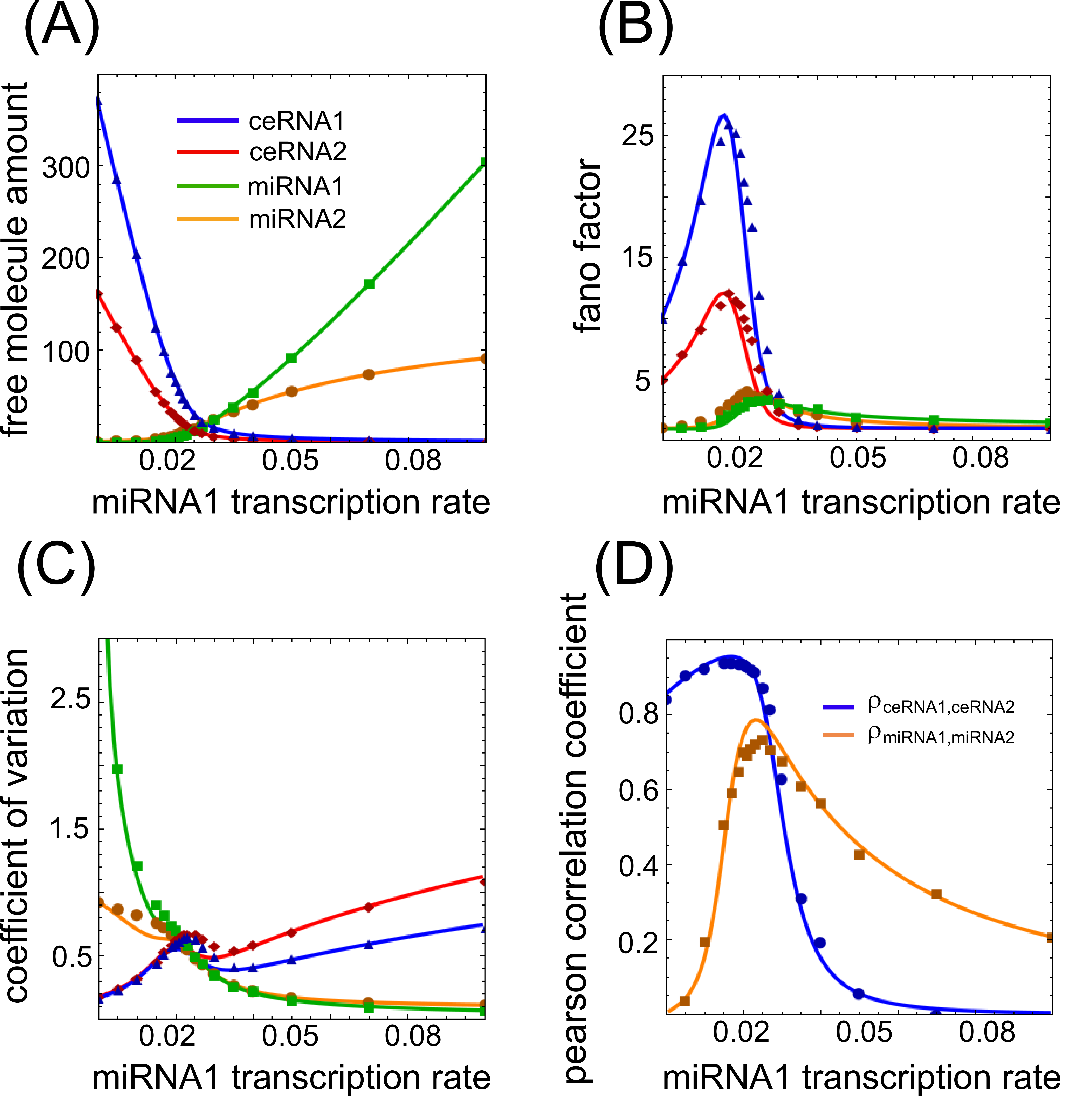}
\caption{{\it Threshold, noise and Pearson's correlation coefficient varying miRNA transcription rate.}(A-C) Steady state value for means, Fano factors and coefficients of
variation for each free molecular species in a system of two miRNAs
(miRNA1 and miRNA2, green and orange lines respectively) interacting
with two ceRNAs (ceRNA1 and ceRNA2, blue and red lines respectively)
varying the concentration of miRNA1. In proximity to the threshold the
system shows hypersensitivity to changes in the control parameter
(miRNA1 transcription rate), captured by a maximum in the Fano factors
(panel B). For the same values of miRNA1 transcription rate, the local
maximum in the coefficients of variation (panel C) is the fingerprint
of bimodal distributions in the number of molecules for each molecular
species.  (D) Pearson's coefficients between the two miRNAs (orange
line) and the two ceRNAs (blue line). The two lines show a maximum in
proximity to the miRNA1 transcriptiom rate threshold value, meaning
that there is a region of parameters where the fluctuations in the
number of ceRNAs or miRNAs are highly correlated.  Lines are the
results of Gaussian approximation while symbols are Gillespie's
simulations. For panels B,C the line color-code is the same as in
panel A.}
\label{duale_mirna}
\end{figure}

\begin{figure}[h]
\includegraphics[scale=0.2]{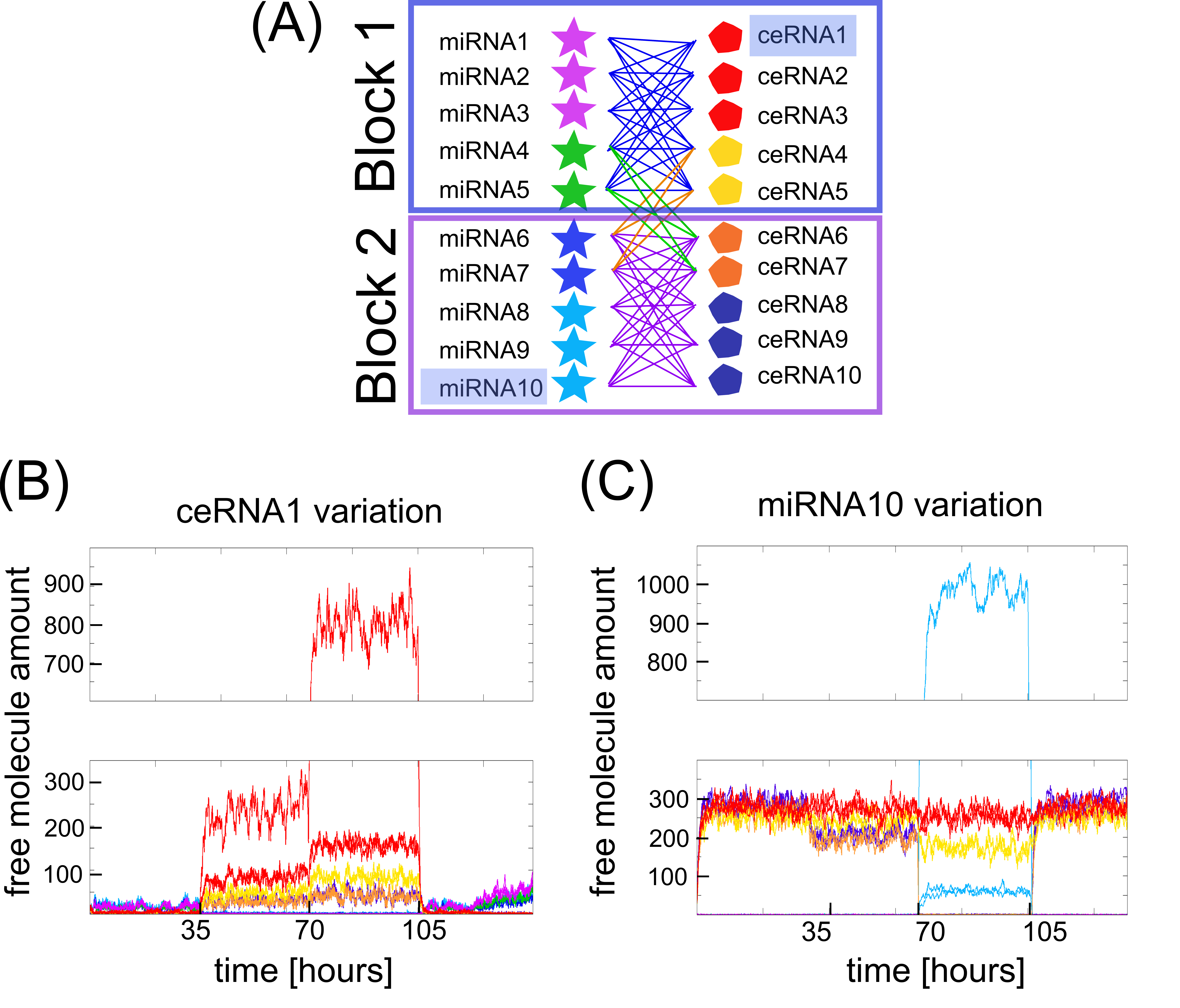}
\caption{{\it Selectivity of miRNA and ceRNA interactions.}(A) Example of a network of ten miRNAs interacting with ten ceRNAs in
blocks. The interaction links are such that we can define two main
blocks (block 1 and block2) of strongly interacting miRNAs-ceRNAs
connected by two common miRNAs (miRNA 4 and 5 in block 1, miRNA 6 and
7 in block 2) and ceRNAs (ceRNA 4 and 5 in block 1 and ceRNA 6 and 7
in block 2). Panels (B,C) show an example of dynamics of such
network. Varying ceRNA1 (panel B) or miRNA10 (panel C) transcription
rate during time (every 35 hours in the example, but the time is
arbitrary) has a differentiated effect on the other ceRNAs and miRNAs
present in the all network. The color-code for lines in panels B and C
follows the color of miRNAs and ceRNAs in panel A.}
\label{select}
\end{figure}

\begin{figure}[h]
\includegraphics[width=1\textwidth]{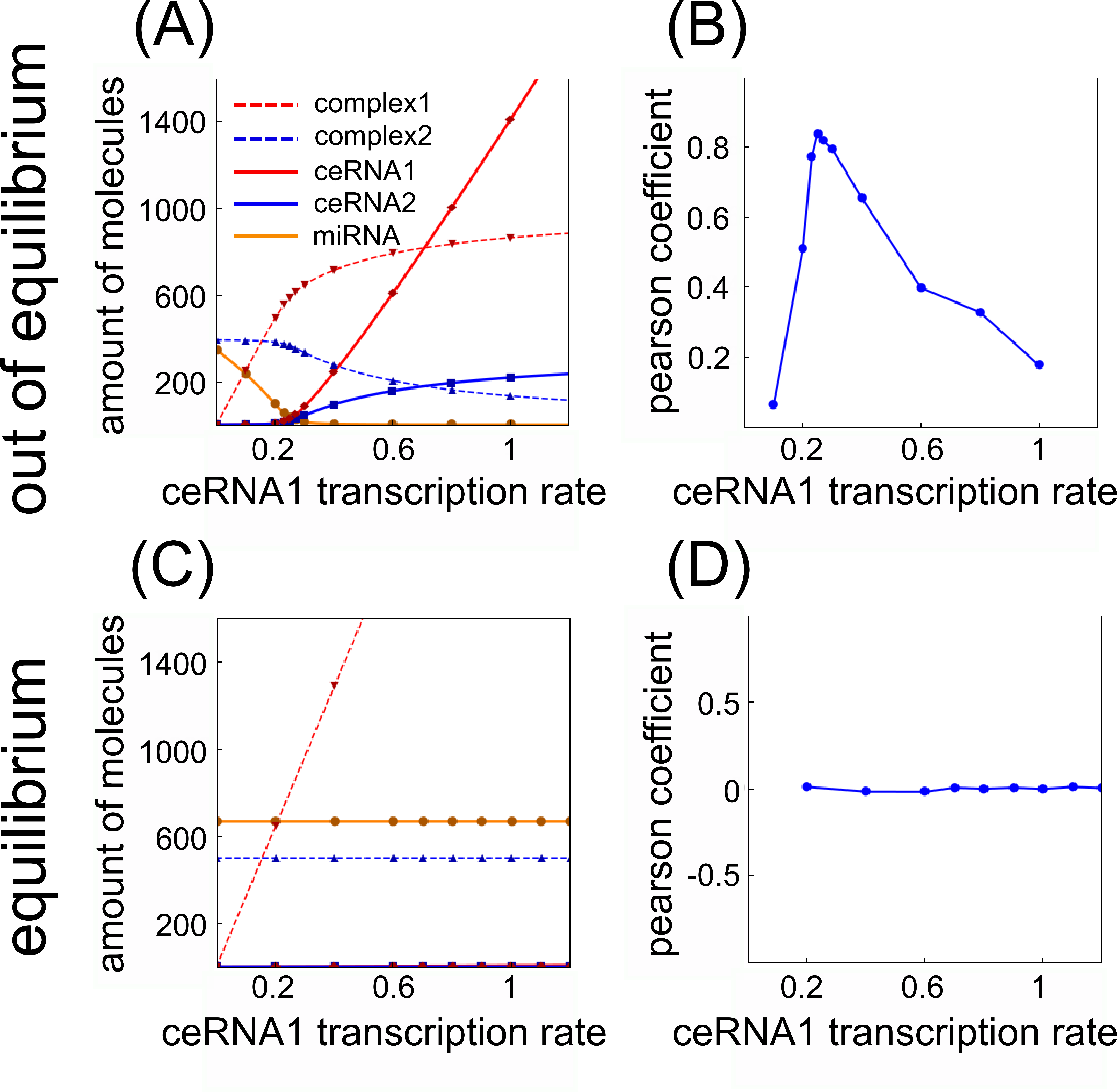}
\caption{{\it Threshold effect in a miRNA-target catalytic interaction.}
  Example of a system of one miRNA interacting with two ceRNAs with
  cataliticity parameter $\alpha = 0$. The threshold effect is
  possible only if the system is out of equilibrium (A). Numerical
  integration of Equation~(1) in Supplementary Materials leads to time
  evolution of each molecular species for a given set of
  parameters. In panels A,C we plot ``pictures" of the evolving system
  at different time $t$ (panel A $t=10^3$, panel C $t=10^6$) as a
  function of ceRNA1 transcription rate. When t is smaller than the
  time complexes need to reach their steady state a threshold effect
  is observed. In panels B,D we plot the corresponding Pearson's
  coefficient profiles. Where the threshold effect is present (panel
  A), a peak in the Pearson's coefficient is also observed.}
\label{ooe}
\end{figure}

\begin{figure}[h]
\includegraphics[width=1\textwidth]{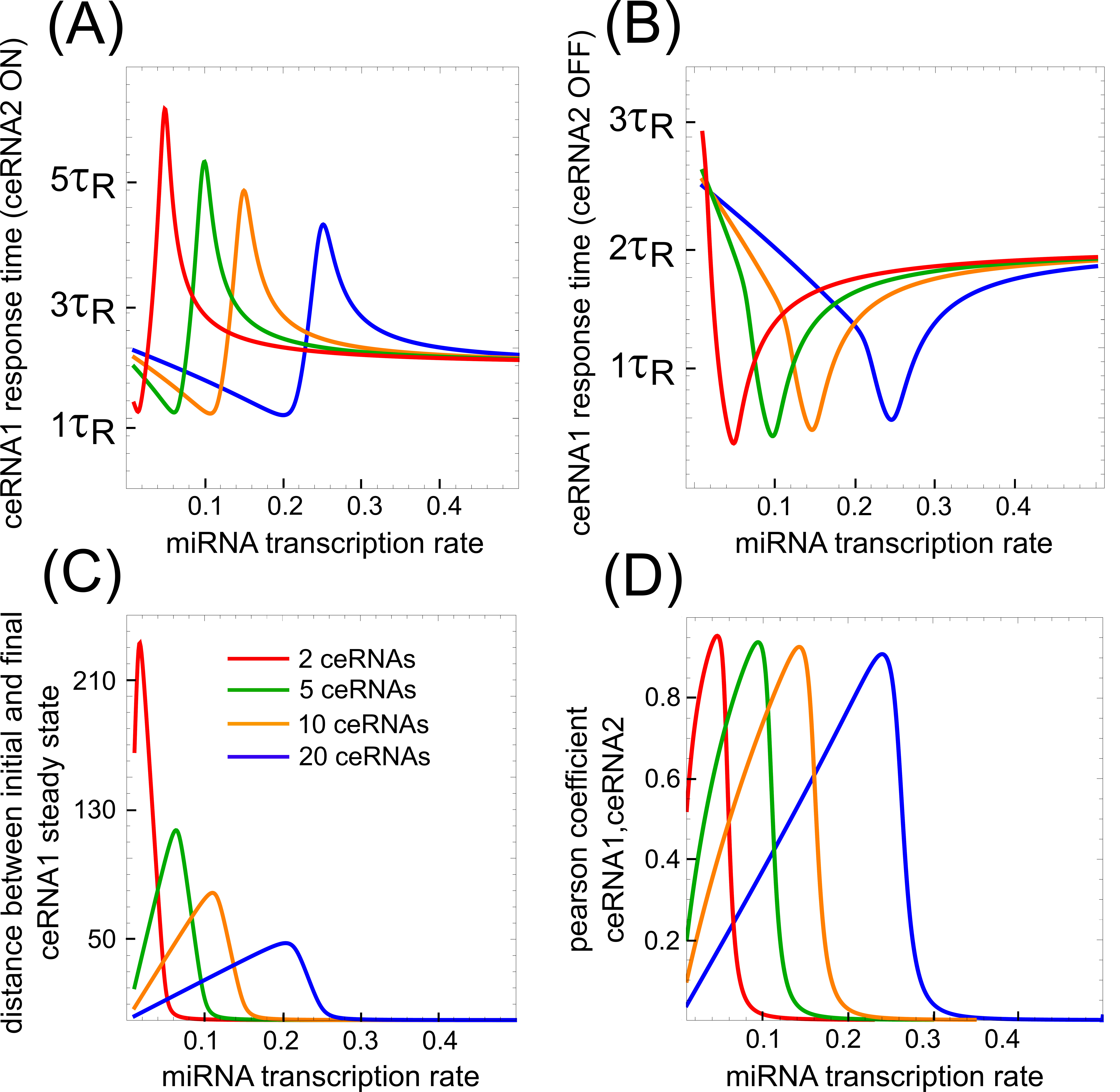}
\caption{{\it Response times upon one ceRNA perturbation.}Increasing miRNA transcription rate ceRNA1 shows a maximum and a
minimum in its response times upon switching on or off ceRNA2
transcription respectively (panel A and B). The maximum (minimum) is
located near the threshold, where ceRNA1 initial value (that is its
values before switching on (off) ceRNA2) is near to the steady state
it will reach upon switching on (off) ceRNA2 (panel C) but also the
more sensitive to ceRNA2 variation (look at the maximum in the
Pearson's correlation coefficient between ceRNA1 and ceRNA2 in panel
D). Different color lines correspond to different numbers of ceRNAs in
competition for the same miRNA. The qualitative trend for response
times and Pearson's correlation coefficient is robust with respect to
increasing such number.}
\label{tempi}
\end{figure}

In analogy with Figure \ref{schema}B, for each gene belonging to the
miRNA-target network in Figure \ref{schema}A we consider the key steps
of transcription, degradation and titrative interaction among
transcripts.  Thus, the system is described by $M+N$ variables ($M$
miRNAs $S_i$ and $N$ target mRNAs $R_j$ transcribed from $M+N$
different genes) and the probability of finding in a cell exactly
${\bf R,S}:=S_1,\dots,S_M,R_1,\dots,R_N$ molecules at time $t$
satisfies the following master equation:
\begin{eqnarray}
\label{master}
\partial_t P &=&  \sum_{i=1}^{M} k_{S_i} (P_{S_i-1} - P) + \sum_{j=1}^{N} k_{R_j} (P_{R_j-1} - P) + \\ \nonumber
&+& \sum_{i=1}^{M} g_{S_i} ((S_i + 1) P_{S_i+1} - S_i P) + \sum_{j=1}^{N} g_{R_j} ((R_j + 1) P_{R_j+1} - R_j P) + \\ \nonumber
&+& \alpha \sum_{i=1}^{M} \sum_{j=1}^{N} g_{ij} ((S_i+1)(R_j+1)P_{S_i+1,R_j+1} - S_i R_j P) + \\ \nonumber
&+& (1-\alpha) \sum_{i=1}^{M} \sum_{j=1}^{N} g_{ij} S_i ((R_j+1) P_{R_j+1} - R_j P) \; ,
\end{eqnarray} 
with $P = P_{X_1, \dots, X_k, \dots , X_{M+N}}$ and $P_{X_k \pm 1} =
P_{X_1, \dots , X_k \pm 1,\dots, X_{M+N}}$.  In Equation \ref{master}
$k_{S_i}$ and $k_{R_j}$ are transcription rates and $g_{S_i}$ and
$g_{R_j}$ degradation rates for the $i$-th miRNA and the $j-$th target
mRNA respectively. $g_{ij}$ is the effective association rate for
miRNA $S_i$ and its target $R_j$. $\alpha$ is the catalyticity
parameter described above.

By defining the generating function,
\begin{equation}
F({\bf{z,q }}| t ) = \sum_{{\bf S,R}} \prod_{i=1}^M \prod_{j=1}^N
z_i^{S_i} q_j^{R_j} P_{{\bf R,S}} \; ,
\label{generating}
\end{equation}
where ${\bf z,q} := z_1,\dots,z_M,q_1,\dots,q_N$, we can convert
Equation \ref{master} into the following second-order partial
differential equation:
\begin{equation}
\label{genfun}
\partial_t F({\bf{z,q }} | t ) = {\cal H}({\bf z,q}) F({\bf{z,q }}| t)  
\end{equation}
where the operator ${\cal H}({\bf z,q})$ is defined as:

\begin{eqnarray}
{\cal H} ({\bf z,q}) &=& \sum_{i=1}^{M} k_{S_i} (z_i - 1) +
\sum_{j=1}^{N} k_{R_j} (q_j - 1) + \\ \nonumber &+& \sum_{i=1}^{M}
g_{S_i} (\partial_{z_i} - z_i \partial_{z_i} ) + \sum_{j=1}^{N}
g_{R_j} (\partial_{q_j} - q_j \partial_{q_j} ) + \\ \nonumber &+&
\alpha \sum_{i=1}^{M} \sum_{j=1}^{N} g_{ij} (\partial^2_{z_i,q_j} -
z_i q_j \partial^2_{z_i,q_j} ) + (1-\alpha)
\sum_{i=1}^{M} \sum_{j=1}^{N} g_{ij} z_i (\partial^2_{z_i,q_j} - q_j
\partial^2_{z_i,q_j} )\; .
\label{genfuncH}
\end{eqnarray}
\\
The moment generating function has the following properties:
\begin{eqnarray}
F({\bf z =1, q =1}) &=& 1 \; , \\ \nonumber
\partial_{z_i} F |_{\bf z =1, q =1} &=& \langle S_i\rangle \; , \\ \nonumber
\partial_{q_j} F  |_{\bf z =1, q =1} &=& \langle R_j\rangle \; , \\ \nonumber
\partial_{z_i}^{2} F  |_{\bf z =1, q =1} &=& \langle S_{i}^{2}\rangle - \langle S_i\rangle \; , \\ \nonumber
\partial_{q_j}^{2} F  |_{\bf z =1, q =1} &=& \langle R_{j}^{2}\rangle - \langle R_j\rangle
\; , \\
\partial_{z_i,q_j}^{2} F  |_{\bf z =1, q =1} &=&  \langle S_i R_j \rangle \; .
\label{property}
\end{eqnarray}
Considering higher order derivatives in Equation \ref{genfun} at
steady state ($\partial_t F = 0$), and assuming that all derivatives
are computed in ${\bf z=1,q=1}$, we find:
\begin{eqnarray}
\label{moments}
\langle S_i\rangle &=& \partial_{z_i} F = 
\frac{k_{S_i} - \alpha \sum_{j=1}^{N} g_{ij} \partial^2_{z_i,q_j} F}{g_{S_i}}\; , \\ \nonumber
\langle R_j\rangle &=& \partial_{q_j} F = 
\frac{k_{R_j} - \sum_{i=1}^{M} g_{ij} \partial^2_{z_i,q_j} F}{g_{R_j}}\; , \\ \nonumber
\langle S_{i}^2\rangle &=& \partial_{z_i}^{2} F + \partial_{z_i} F = 
\frac{k_{S_i} (1 + \partial_{z_i} F) - \alpha \sum_{j=1}^{N} g_{ij} (\partial_{z_i^2,q_j}^3  F + \partial^2_{z_i,q_j} F)}{g_{S_i}} \; , \\ \nonumber 
\langle R_{j}^2\rangle &=& \partial_{q_j}^{2} F + \partial_{q_j} F = 
\frac{k_{R_j} (1 + \partial_{q_j} F) - \sum_{i=1}^{M} g_{ij} (\partial^3_{z_i,q_j^2}  F + \partial^2_{z_i,q_j} F)}{g_{R_j}} \; , \\ \nonumber 
\langle S_i R_j\rangle &=& \frac{k_{R_j} \partial_{z_i} F + k_{S_i}
  \partial_{q_j} F - \sum_{l=1}^{M} g_{lj} \partial^3_{z_i.z_l,q_j} F
  - \alpha \sum_{l=1}^{N} g_{il} \partial^3_{z_i,q_j,q_l} F}{g_{ij} + g_{S_i} + g_{R_j}} \; , \\ \nonumber 
\mbox{etc...} \; .
\end{eqnarray}
\\
The moment-generating function defined in Equation \ref{genfun} is
unfortunately too complicated to be computed analytically even at
steady state, as all moments depend on higher ones and the system is
not closed, as shown in Equation \ref{moments}. In the following we
will present a series of increasingly accurate approximations for
analyzing it.


\subsection*{Independent molecular-species approximation}

As a first step for determining analytically the behavior of the
system, we will assume that the probability distribution $P$ is
factorized:
\begin{equation}
\label{eq:independent}
P^{\mathrm{ind}}({\bf R,S}) := \prod_{i=1}^M P^S_i(S_i)  \prod_{j=1}^N P^R_j(R_j)
\end{equation}
Under this assumption it turns out that the steady state solution for
the $P_i^S(S_i)$, and $ P_j^R(R_j)$ are Poisson distributions whose
mean value can be expressed solving the following second order system
of equations,
\begin{eqnarray}
\langle S_i \rangle_{\mathrm{ind}} &=& \frac{k_{S_i} - \alpha
\langle S_i\rangle_ {\mathrm{ind}} \sum_{j=1}^{N} g_{ij} \langle R_j
\rangle_ {\mathrm{ind}} }{g_{S_i}} \;\;\;\;\; 1 \leq i\leq M \\ \nonumber
\langle R_j \rangle_{\mathrm{ind}} &=& \frac{k_{R_j} - 
\langle R_j\rangle_ {\mathrm{ind}} \sum_{i=1}^{M}g_{ij} \langle S_i
\rangle_ {\mathrm{ind}} }{g_{R_j}}\;\;\;\;\;\;\; 1 \leq j \leq N\; .
\label{indep}
\end{eqnarray}

Analytic solutions for the system of equations \ref{indep} can be
easily written in the case $g_{R_j} = g_R$, $g_{S_i} = g_S$ and
$g_{ij} = g$ for all $R_j$ and $S_i$:
\begin{eqnarray}
\langle S_q \rangle_{\mathrm{ind}} &=& \frac{k_{S_q}}{2g_S \sum_{i=1}^M k_{S_i}}
\left(k_{S_q} +  \sum_{i \neq q}^M k_{S_i} -  \alpha \sum_{j=1}^N k_{R_j} - \frac{g_R g_S - \sqrt{A}}{g}\right)\\
\langle R_p \rangle_{\mathrm{ind}} &=& \frac{k_{R_p}}{2 g_R \sum_{j=1}^{N} k_{R_j}}
\left(
k_{R_p} + \sum_{j \neq p}^{N}k_{R_j} - \sum_{i=1}^{M}k_{S_i} - \frac{g_R g_S - \sqrt{A}}{\alpha g}
\right)\quad,\nonumber
\label{indep2}
\end{eqnarray}

with $A = 4 g g_S g_R \alpha \sum_{j=1}^{N}k_{R_j} + (g_R g_S +
g(\sum_{i=1}^{M} k_{S_i} - \alpha \sum_{j=1}^{N} k_{R_j}))^2$ .  In
the more general and biologically relevant case of different molecules
half-lives and complex affinities $g_{ij}$, solutions can still be
found, but they turn out to be too complex and long to be reported
here.


\subsection*{Locating the threshold}

The simplest way to locate the threshold is to solve the system of
equations \ref{indep} in the limit of strong interaction miRNA-target
(high $g_{ij}$) thus finding:

\begin{eqnarray}
\label{soglia}
\langle S_i \rangle_{\mathrm{ind,ss}} &\rightarrow& \left\{ 
\begin{array}{l l} \frac{ k_{S_i} - \alpha \sum_{j=1}^{N} k_{R_j}}{g_{S_i}} & \quad \mbox{if} \quad \alpha \sum_{j=1}^{N} k_{R_j} < \sum_{i=1}^{M} k_{S_i} \\ 
  0 & \quad \mbox{otherwise}\\
  \end{array} \right. \\ \nonumber
\langle R_j \rangle_{\mathrm{ind,ss}} &\rightarrow& \left\{ 
\begin{array}{l l} \frac{k_{R_j} - \sum_{i=1}^{M}k_{R_j} k_{S_i}/(\alpha \sum_{j=1}^{N} k_{R_j})} {g_{R_j}} & \quad \mbox{if} \quad \alpha \sum_{j=1}^{N} k_{R_j}\geq \sum_{i=1}^{M} k_{S_i} \\
0 & \quad  \mbox{otherwise}\\
\end{array}\right. 
\end{eqnarray}

The threshold position is determined by the relative amount of miRNAs
and their targets (see Equation~\ref{soglia}).  For fixed $k_{R_j}$
and $k_{S_i}$, with $j = \{1,...,q-1,q+1,...,N\}$ and $i =
\{1,...,M\}$, the threshold is set by $k_{R_j}$ and by all miRNA
transcription rates $k_{S_i}$.  Thus, as long as the q-th mRNA target
transcription rate $k_{R_q}$ is below its threshold level $k_{R_q}^{*}
= (\sum_{i=1}^{M} k_{S_i}-\alpha \sum_{j \neq q}^{N} k_{R_j})/\alpha $
all targets are bound in complexes and their free molecule amount is
very low (while miRNAs are expressed), or, in other terms, the
threshold is located at near-equimolar concentration of the different
chemical species.

Increasing $k_{R_q}$ beyond its threshold results in the expression of
all the other targets (while miRNAs will be all bound in complexes),
see Figure~\ref{duale_cerna}A.

Within the independent chemical species approximation in
Equation~\ref{eq:independent} the Fano factor (noise index $f_\langle
X \rangle = \sigma_{\langle X \rangle}^2 / \langle X\rangle$) for each molecular
species is 1.  The factorized approximation is good enough in showing
the threshold effect, but fails in determining correlations among
molecular species (see symbols, which are the results of Gillespie's
simulations, in Figures \ref{duale_cerna}A and
\ref{duale_mirna}A).

\subsection*{Gaussian Approximation}

The simplest approximation beyond mean-field is a Gaussian one. We
denote  $\vec X = (X_1,\dots,X_{N+M}) :=
(R_1,\dots,R_N,S_1,\dots,S_M)$. The approximation assumes that $\vec
X$ is distributed as a multivariate Gauss:
\begin{equation}
\label{eq:gauss}
P(\vec X)  =  \frac{\exp\left[ -\frac12 (\vec X - \vec \mu)^T C^{-1} (\vec X - \vec \mu)\right]}{\sqrt{(2 \pi)^{N+M}\mathrm{det}(C)}}\,\,\,\,,
\end{equation} 
where the covariance matrix $C$ has coordinates $c_{ij}:= E(X_iX_j) -
E(X_i) E(X_j)$, the vector $\vec \mu $ has coordinates $\mu_i :=
E(X_i)$, and the expectation value $E(\cdot)$ is with respect to the
Gaussian measure $P$ defined in Equation~\ref{eq:gauss}. All moments
of a Gaussian multivariate measure can be expressed in terms of
$\mu_i$ and $c_{ij}$. Therefore the moments derived from the generating
function in Equation~\ref{moments} can be expressed in terms of
$\mu_i$ and $c_{ij}$. In the Supplementary Material we describe in
details the computation of the specific $N=M=2$ case, and we compare
the performance of the Gaussian approximation with the linear-noise
approximation.


\section*{Acknowledgments}
While completing this manuscript we learned that M. Figliuzzi,
E. Marinari, and A. De Martino have independently studied the same
problem, reporting results which are consistent with those obtained
here.

We thank Michele Caselle, Enzo Marinari, Paolo Provero, Andrea De
Martino, Luca Dall'Asta, Carlo Baldassi, Matteo Osella, Marco Zamparo,
and Matteo Figliuzzi, for interesting discussions about technical
aspects of stochastic modeling. We are indebted with Pier Paolo
Pandolfi, Yvonne Tay, Florian Karreth, and Riccardo Taulli for many
illuminating discussions about the experimental strategies for
validating the model, and Terence Hwa for pointing us a relevant
bibliographic reference on the subject. RZ acknowledges support from
the ERC Grant No. OPTINF 267915.







  


\clearpage
\newpage 
\appendix

{\huge
\begin{center}
{\bf Supplementary Material}
\end{center}
}

\section{Generalized mean-field equation with explicit complexes}

\begin{figure}[h]
\centering
\includegraphics[scale=0.2]{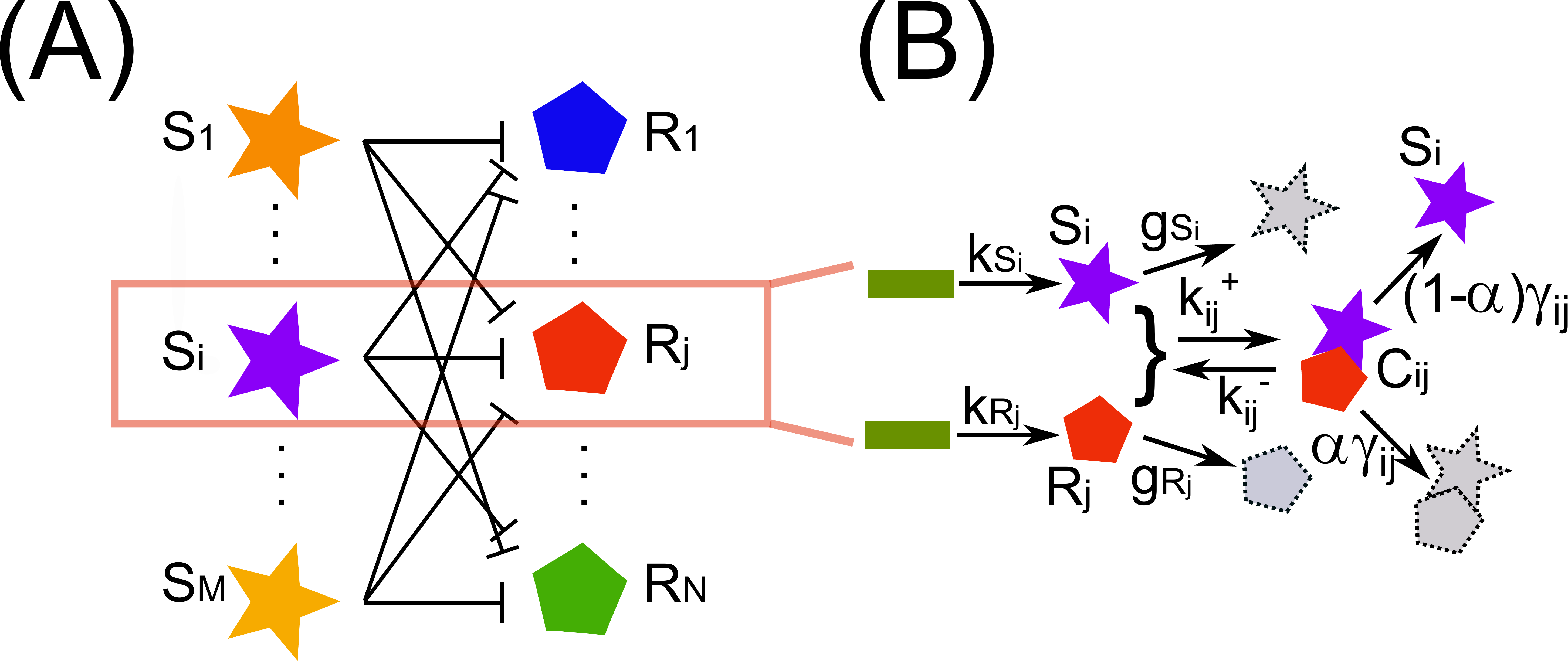}
\caption{{\it Representation of a generic miRNA-target interaction
    network.} (A) Simplified picture of a miRNA-ceRNA interaction
  network. (B) For each miRNA ($S_i$) and ceRNA ($R_j$) present in the
  network we consider the main steps of transcription (rates $k_{S_i}$
  and $k_{R_j}$ respectively) and degradation (rates $g_{S_i}$ and
  $g_{R_j}$ respectively) plus a titrative interaction between miRNA
  and ceRNA. miRNA and ceRNA can therefore form a complex $C_{ij}$
  with association rate $k_{ij}^+$and dissociation rate $k_{ij}^-$.
  The complex can then degrade with rate $\gamma_{ij}$. The parameter
  $\alpha$ (the catalyticity parameter) tells which is the probability
  a miRNA is recycled after having interact with one of its targets.}
\label{schema_SI}
\end{figure}

We describe the general case of $N$ different target mRNAs interacting
with $M$ different miRNAs. The action of a miRNA on its target has the
following characteristics: each miRNA molecule can constitute a
complex with a target molecule and then can be eventually released.
The molecular species are: free miRNAs ($S_i$), free mRNAs ($R_j$),
complexes $C_{ij}$ of miRNA $S_i$ with mRNA $R_j$. Each gene can be
transcribed with rate $k_{\{R_i,S_j\}}$, degraded with rate
$g_{\{R_i,S_j\}}$. Complexes $C_{ij}$ associate with rate $k_{ij}^+$
and dissociate with rate $k_{ij}^-$. Each complex eventually degrade
with rate $\gamma_{ij}$. A schema of such network is represented in
Figure~\ref{schema_SI}.  The mean-field equations thus reads:
\begin{eqnarray}
\frac{dR_i}{dt} &=& k_{R_i} - g_{R_i} S_j + \sum_{j=1}^M \left(  
-k_{ij}^+ S_jR_i + k_{ij}^- C_{ij} \right) \nonumber\\
\frac{dS_j}{dt} &=& k_{S_j} - g_{S_j} S_j + \sum_{i=1}^N \left(  
-k_{ij}^+ S_iR_j + k_{ij}^- C_{ij} +
(1-\alpha)\gamma_{ij}C_{ij}\right)\nonumber\\
\frac{dC_{ij}}{dt} &=& k_{ij}^+ R_i S_j - (k_{ij}^- +  \gamma_{ij}) C_{ij} 
\label{meanField}
\end{eqnarray}
with $j\in\{i,\dots,M\}$ and $i\in\{ 1,\dots,N\}$.  Assuming that
complexes reach the equilibrium faster than the other molecular
species, we can simplify the system \ref{meanField} to the following one:

\begin{eqnarray}
\label{meanfield}
\frac{dS_i}{dt} &=& k_{S_i} - g_{S_i} S_i - \alpha g_{ij} S_i R_j \\ \nonumber
\frac{dR_j}{dt} &=& k_{R_j} - g_{R_j} R_j - g_{ij} S_i R_j \; ,
\end{eqnarray}  
with $g_{ij} = \frac{k_{ij}^{+} \gamma_{ij}}{k_{ij}^{-} + \gamma_{ij}}$.

\section{Generalized master equation with explicit complexes}

The master equation corresponding to Equation \ref{meanField} reads:
\begin{eqnarray}
\partial_t P &=&  \sum_{i=1}^{M} k_{S_i} (P_{S_i-1,R_j,C_{ij}} - P_{S_i,R_j,C_{ij}}) + \sum_{j=1}^{N} k_{R_j} (P_{S_i,R_j-1,C_{ij}} - P_{S_i,R_j,C_{ij}}) + \\ \nonumber
&+& \sum_{i=1}^{M} g_{S_i} ((S_i + 1) P_{S_i+1,R_j,C_{ij}} - S_i P_{S_i,R_j,C_{ij}}) + \sum_{j=1}^{N} g_{R_j} ((R_j + 1) P_{S_i,R_j+1,C_{ij}} - R_j P_{S_i,R_j,C_{ij}}) + \\ \nonumber
&+& \sum_{i=1}^{M} \sum_{j=1}^{N} k_{ij}^{+} ((S_i+1)(R_j+1) P_{S_i+1,R_j+1,C_{ij}-1} - S_i R_j P_{S_i,R_j,C_{ij}}) + \\ \nonumber
&+& \sum_{i=1}^{M} \sum_{j=1}^{N} k_{ij}^{-} ((C_{ij}+1) P_{S_i-1,R_j-1,C_{ij}+1} - C_{ij} P_{S_i,R_j,C_{ij}}) + \\ \nonumber
&+& \alpha \sum_{i=1}^{M} \sum_{j=1}^{N} \gamma_{ij} ((C_{ij}+1) P_{S_i,R_j,C_{ij}+1} - C_{ij} P_{S_i,R_j,C_{ij}}) + \\ \nonumber
&+& (1-\alpha) \sum_{i=1}^{M} \sum_{j=1}^{N} \gamma_{ij} ((C_{ij}+1) P_{S_i-1,R_j,C_{ij}+1} - C_{ij} P_{S_i,R_j,C_{ij}}) \; ,
\label{masterCompleta}
\end{eqnarray}


\section{Gaussian Approximation}

We work here in some details the explicit computation for the Gaussian
approximation in the specific case of 2 microRNAs ($S_1,S_2$) and 2
ceRNAs ($R_1,R_2$). Denoting with $\partial^{l+m}_{z_i^l,q_j^m}F :=
\partial^{l+m}_{z_i^l,q_j^m}F |_{\mathbf{z,q} = (1,\dots,1)} $, at steady state
the system of equation reads:
\begin{eqnarray}
  \partial_{z_1}F &=& \frac{k_{z_1} - \alpha g ( \partial^2_{z_1,q_1} F +
    \partial^2_{z_1,q_2}F)}{g_S} \nonumber \\
  \partial_{z_2}F &=&  \frac{k_{z_2} - \alpha g ( \partial^2_{z_2,q_1} F +
    \partial^2_{z_2,q_2}F)}{g_S} \nonumber\\ 
  \partial_{q_1}F &=& \frac{k_{q_1} -  g ( \partial^2_{z_1,q_1} F +
    \partial^2_{z_2,q_1}F)}{g_R} \nonumber \\
  \partial_{q_2}F &=&  \frac{k_{q_2} - g ( \partial^2_{z_1,q_2} F +
    \partial^2_{z_2,q_2}F)}{g_R}\nonumber \\
  \partial^2_{z_1^2} F &=& \frac{ k_{z_1} \partial_{z_1}F - \alpha g (
    \partial^3_{z_1^2,q_1}F + \partial^3_{z_1^2,q_2}F)}{ g_S}
  \nonumber \\
  \partial^2_{z_1,z_2} F &=& \frac{ k_{z_2} \partial_{z_1}F  +
    k_{z_1}\partial_{z_2}F - 2\alpha g ( \partial^3_{z_1,z_2,q_1} F +
    \partial^3_{z_1,z_2,q_2}F)}{2g_S}\nonumber \\
  \partial^2_{z_1,q_1} F &=& \frac{ k_{q_1} \partial_{z_1}F  +
    k_{z_1}\partial_{q_1}F -  g ( \partial^3_{z_1^2,q_1} F +
    \partial^3_{z_1,z_2,q_1}F + \alpha \partial^3_{z_1,q_1^2}F + \alpha
    \partial^3_{z_2,q_1,q_2}F )}{g + g_R + g_S} \nonumber \\
   \partial^2_{z_1,q_2} F &=& \frac{ k_{q_2} \partial_{z_1}F  +
     k_{z_1}\partial_{q_2}F - g ( \partial^3_{z_1^2,q_2} F +
    \partial^3_{z_1,z_2,q_2}F + \alpha \partial^3_{z_1,q_1,q_2}F + 
    \partial^3_{z_2,R^2_2}F )}{g + g_R + g_S} \nonumber \\
   \partial^2_{z_2^2} F &=& \frac{ k_{z_2} \partial_{z_2}F - \alpha g (
    \partial^3_{z_2^2,q_1}F + \partial^3_{z_2^2,q_2}F)}{ g_S} \nonumber\\
   \partial^2_{z_2,q_1} F &=& \frac{ k_{q_1} \partial_{z_2}F  +
     k_{z_2}\partial_{q_1}F - g ( \partial^3_{z_1,z_2,q_1} F +
    \partial^3_{z_2^2,q_1}F + \alpha \partial^3_{z_2,q_1^2}F + 
    \alpha \partial^3_{z_2,q_1,q_2}F )}{g + g_R + g_S} \nonumber \\
   \partial^2_{z_2,q_2} F &=& \frac{ k_{q_2} \partial_{z_2}F  +
     k_{z_2}\partial_{q_2}F - g ( \partial^3_{z_1,z_2,q_1} F +
    \partial^3_{z_2^2,q_1}F + \alpha \partial^3_{z_2,q_1^2}F + 
    \alpha \partial^3_{z_2,q_1,q_2}F )}{g + g_R + g_S} \nonumber \\
   \partial^2_{q_1^2} F &=& \frac{ k_{q_1} \partial_{q_1}F -  g (
    \partial^3_{z_1,q_1^2}F + \partial^3_{z_2,q_1^1}F)}{ g_R} \nonumber\\
   \partial^2_{q_1,q_2} F &=& \frac{ k_{q_2} \partial_{q_1}F  +
     k_{q_1}\partial_{q_2}F - 2 g ( \partial^3_{z_1,q_1,q_2} F +
     \partial^3_{z_2,q_1,q_2}F)}{g + g_R + g_S} \nonumber \\
   \partial^2_{q_2^2} F &=& \frac{ k_{q_2} \partial_{q_2}F -  g (
    \partial^3_{z_1,q_2^2}F + \partial^3_{z_2,q_2^2}F)}{ g_R}
\label{eq:gauss2x2}
\end{eqnarray}

Recalling that within the Gaussian approximation the partial
derivatives of the third order can be expressed in terms of that of
lower order:
\begin{eqnarray}
\partial^3_{z_1^2, z_2} F &=& 
(\partial^2_{z_1^2}F + \partial_{z_1}F)\partial_{z_2}F + 2
\partial_{z_1}F \partial^2_{z_1,z_2}F - 2 (\partial_{z_1}F)^2
\partial_{z_2}F-\partial^2_{z_1,z_2}F\nonumber\\
\partial^3_{z_1^2, q_1} F &=& 
(\partial^2_{z_1^2}F + \partial_{z_1}F)\partial_{q_1}F + 2
\partial_{z_1}F \partial^2_{z_1,q_1}F - 2 (\partial_{z_1}F)^2
\partial_{q_1}F-\partial^2_{z_1,q_1}F\nonumber\\
\partial^3_{z_1^2, q_2} F &=& 
(\partial^2_{z_1^2}F + \partial_{z_1}F)\partial_{q_2}F + 2
\partial_{z_1}F \partial^2_{z_1,q_2}F - 2 (\partial_{z_1}F)^2
\partial_{q_2}F-\partial^2_{z_1,q_2}F\nonumber\\
\partial^3_{z_1, z_2^2} F &=& 
(\partial^2_{z_2^2}F + \partial_{z_2}F)\partial_{z_1}F + 2
\partial_{z_2}F \partial^2_{z_1,z_2}F - 2 (\partial_{z_2}F)^2
\partial_{z_1}F-\partial^2_{z_1,z_2}F\nonumber\\
\partial^3_{z_1, q_1^2} F &=& 
(\partial^2_{q_1^2}F + \partial_{q_1}F)\partial_{z_1}F + 2
\partial_{q_1}F \partial^2_{z_1,q_1}F - 2 (\partial_{q_1}F)^2
\partial_{z_1}F-\partial^2_{z_1,q_1}F\nonumber\\
\partial^3_{z_1, q_2^2} F &=& 
(\partial^2_{q_2^2}F + \partial_{q_2}F)\partial_{z_1}F + 2
\partial_{q_2}F \partial^2_{z_1,q_2}F - 2 (\partial_{q_2}F)^2
\partial_{z_1}F-\partial^2_{z_1,q_2}F\nonumber\\
\partial^3_{z_2^2, q_1} F &=& 
(\partial^2_{z_2^2}F + \partial_{z_2}F)\partial_{q_1}F + 2
\partial_{z_2}F \partial^2_{z_2,q_1}F - 2 (\partial_{z_2}F)^2
\partial_{z_1}F-\partial^2_{z_1,q_1}F\nonumber\\
\partial^3_{z_2^2, q_2} F &=& 
(\partial^2_{z_2^2}F + \partial_{z_2}F)\partial_{q_2}F + 2
\partial_{z_2}F \partial^2_{z_2,q_2}F - 2 (\partial_{z_2}F)^2
\partial_{z_1}F-\partial^2_{z_1,q_2}F\nonumber\\
\partial^3_{z_2, q_1^2} F &=& 
(\partial^2_{q_1^2}F + \partial_{q_1}F)\partial_{z_2}F + 2
\partial_{q_1}F \partial^2_{z_2,q_1}F - 2 (\partial_{q_1}F)^2
\partial_{z_2}F-\partial^2_{z_2,q_1}F\nonumber\\
\partial^3_{z_2, q_2^2} F &=& 
(\partial^2_{q_2^2}F + \partial_{q_2}F)\partial_{z_2}F + 2
\partial_{q_2}F \partial^2_{z_2,q_2}F - 2 (\partial_{q_2}F)^2
\partial_{z_2}F-\partial^2_{z_2,q_2}F\nonumber\\
\partial^3_{z_1, z_2,q_1} F &=& 
\partial^2_{z_1,z_2}\partial_{q_1}F +
\partial^2_{z_2,q_1}\partial_{z_1}F +
\partial^2_{z_1,q_1}\partial_{z_2}F - 2
\partial_{z_1}F \partial_{z_2}F \partial_{q_1}F\nonumber\\
\partial^3_{z_1, z_2,q_2} F &=& 
\partial^2_{z_1,z_2}\partial_{q_2}F +
\partial^2_{z_2,q_2}\partial_{z_1}F +
\partial^2_{z_1,q_2}\partial_{z_2}F - 2
\partial_{z_1}F \partial_{z_2}F \partial_{q_2}F\nonumber\\
\partial^3_{z_2, q_1,q_2} F &=& 
\partial^2_{z_2,q_1}\partial_{q_2}F +
\partial^2_{z_2,q_2}\partial_{q_1}F +
\partial^2_{q_1,q_2}\partial_{z_2}F - 2
\partial_{z_2}F \partial_{q_1}F \partial_{q_2}F\nonumber\\
\partial^3_{z_1, q_1,q_2} F &=& 
\partial^2_{z_1,q_1}\partial_{q_2}F +
\partial^2_{z_1,q_2}\partial_{q_1}F +
\partial^2_{q_1,q_1}\partial_{z_1}F - 2
\partial_{z_1}F \partial_{q_1}F \partial_{q_2}F
\label{eq:thirdmom}
\end{eqnarray}
Inserting relations (\ref{eq:thirdmom}) into (\ref{eq:gauss2x2}) we
obtain a closed system of 14  in 14 unknowns. In the general case of a
network of $N$ ceRNAs interacting through $M$ miRNAs we would have a
complete system of $2(N+M) + {N+M \choose 2}$ equations.


\section{Linear noise approximation}

We use the linear noise approximation~\cite{van2007stochastic} in order to
obtain the steady state fluctuation covariance matrix directly from
the macroscopic system.  For a general system of $M$ miRNAs interacting
with $N$ mRNAs and $R$ elementary reactions, we assign to each reaction
$r$ a propensity $f_r$ defined from the probability $\Omega f_r(\psi,
\Omega)\delta t$ that a reaction $r$ occurs in the homogeneous system
volume $\Omega$ in the time interval $\delta t$. $\psi$ is the
concentration vector of the $M+N$ chemical components of the system.
In the macroscopic limit ($\Omega \rightarrow \infty$) the system
dynamics is described by the following $M+N$ ordinary differential
equations,
\begin{eqnarray}
\frac{d \psi_p}{dt} & = & \sum_{r} \nu_{rp} f_r(\psi_1, ..., \psi_{M+N}) \; ,
\label{psi}
\end{eqnarray}

where $\nu_{rp}$ is the $rp-$th element of the stoichiometry
matrix, i.e. it indicates the number of molecules by which a component
$p$ changes when an elementary reaction of type $r$ occurs.

For small enough deviations $\delta \psi = [\delta \psi_1, \delta
  \psi_2, ..., \delta \psi_{M+N}]$ from its steady state, the dynamics
of Equation (\ref{psi}) can be approximated by a system of linear
differential equations, according to $\frac{\delta \psi}{dt} =
{\mathcal A} \delta \psi$, where ${\mathcal A}$ is the Jacobian matrix
with elements
\begin{eqnarray}
a_{pq} & = & \sum_{r=1}^{R} \nu_{rp} (\frac{\partial f_r}{\partial \psi_q})_{\psi_{ss}} \; .
\end{eqnarray} 

The master equation for the probability of having $X = [X_1,X_2, ..., X_p, ..., X_{M+N}]$ molecules in the system at time t is then
\begin{eqnarray}
\frac{dP}{dt}(X,t) & = & \Omega \sum_{r=1}^{R}(\prod_{p=1}^{N} E_{p}^{\nu_{rp}} -1) f_r (X \Omega^{-1},\Omega) P(X,t) \; ,
\end{eqnarray} 
\\
with $E$ being a step operator with property $E_{p}^{\nu_{rp}} g(..., X_p, ...) = g(..., X_p + \nu_{rp}, ...)$.

To obtain the linear noise approximation \cite{van2007stochastic} we
expand the master equation to second order in $\Omega^{-1/2}$ after
substituting each $p$-th component of $X$ with $X_p = \Omega \psi_p +
\Omega^{1/2} x_p$.  $x_p$ is the $p$-th component of a new random
vector $x$ such that the $X_p$ is thus described as a macroscopic term
$\Omega \psi_p$ plus a stochastic term $\Omega^{1/2} x_p$.  We thus
obtain a linear Fokker-Planck equation for the joint probability
distribution $\Pi(x,t)$ of $x$:
\begin{eqnarray}
\frac{d \Pi}{dt}(x,t) & = & -\sum_{p,q} a_{pq} \frac{\partial x_p \Pi}{\partial x_p} + \frac{1}{2} \sum_{p,q} b_{pq} \frac{\partial^2 \Pi}{\partial x_p x_q}\; .
\label{fokker}
\end{eqnarray} 
The matrix elements $a_{pq}$ are given by the Jacobian matrix
${\mathcal A}$, while the elements $b_{pq}$ of the diffusion matrix
${\mathcal B}$ are defined as in \cite{Risken96},
\begin{eqnarray}
b_{pq} & = & \sum_{r=1}^{R} f_r \nu_{rp} \nu_{pq} \; .
\end{eqnarray} 

Generally ${\mathcal A}$ and ${\mathcal B}$ may depend on time, but
here we will restrict our analysis to the steady state case. In this
way, the stationary solution of Equation (\ref{fokker}) is the normal
distribution $N(0,\Xi)$.  $\Xi$, which is the covariance matrix with
elements $\xi_{rp}$, is the solution of the matrix Lyapunov equation:
\begin{eqnarray}
\mathcal A \Xi + \Xi \mathcal A^{T} + \mathcal B = 0
\label{eqC}
\end{eqnarray}

The covariance matrix ${\mathcal C}$ for the deviations in molecule
number ($\delta X_i$) is related to $\Xi$ via the relation ${\mathcal
  C} = \Omega \Xi$.  Thus, in the linear noise approach the expected
value $\langle X_r \rangle$ is approximated by $\Omega \psi_r$ and the true
covariance $\sigma_{rp}^{2}$ by $c_{rp}$.  Then, the expressions for
Pearson's correlation coefficients ($\rho_{X_r,X_p}$), Fano factors
($f_X$) and coefficients of variation ($CV_X$) can be easily derived:
\begin{eqnarray}
\rho_{X_r,X_p} & = & \frac{\sigma_{rp}^2}{\sigma_{rr}\sigma_{pp}} \sim \frac{c_{rp}}{\sqrt{c_{rr} c_{pp}}} = \frac{\xi_{rp}}{\sqrt{\xi_{rr}\xi_{pp}}} \; , \nonumber \\
CV_{X_r} & = & \frac{\sigma_{rr}}{\langle X_r \rangle} \sim \frac{\sqrt{c_{rr}}}{\Omega \psi_r} = \frac{\sqrt{\xi_{rr}}}{\psi_r} \; , \nonumber \\
f_{X_r} & = & \frac{\sigma^{2}_{rr}}{\langle X_r \rangle} \sim \frac{c_{rr}}{\Omega \psi_r} = \frac{\xi_{rr}}{\psi_r} \; .
\end{eqnarray} 

Therefore, thanks to Equation (\ref{eqC}) the matrix ${\mathcal C}$
(and thus the stochastic properties of a system) can be directly
evaluated from macroscopic parameters.

Let's now discuss in details the specific case with two ceRNAs in
interaction with one miRNAs. In such a system, the propensity vector $f$
assumes the following form:
\begin{eqnarray}
f := \{k_{S_1}, S_1 g_{S_1}, g_{11} S_1 R_1, g_{12} S_1 R_2, k_{S_2}, S_2 g_{S_2}, g_{21} S_2 R_1, g_{22} S_2 R_2, k_{R_1}, R_1 g_{R_1}, k_{R_2}, R_2 g_{R_2}\} \; ,
\end{eqnarray}
\\
and the stoichiometry matrix ${\mathcal \nu}$ is given by:
\begin{equation}
{\mathcal \nu} = \left(
\begin{array}{cccc}
1 & 0 & 0 & 0\\
-1 & 0 & 0 & 0\\
-\alpha & 0 & -1 & 0\\
-\alpha & 0 & 0 & -1\\
0 & 1 & 0 & 0\\
0 & -1 & 0 & 0\\
0 & -\alpha & -1 & 0\\
0 & -\alpha & 0 & -1\\
0 & 0 & 1 & 0\\
0 & 0 & -1 & 0\\
0 & 0 & 0 & 1\\
0 & 0 & 0 & -1\\
\end{array}
\right) \; .
\end{equation}

\begin{figure}[h]
\centering
\includegraphics[scale=0.2]{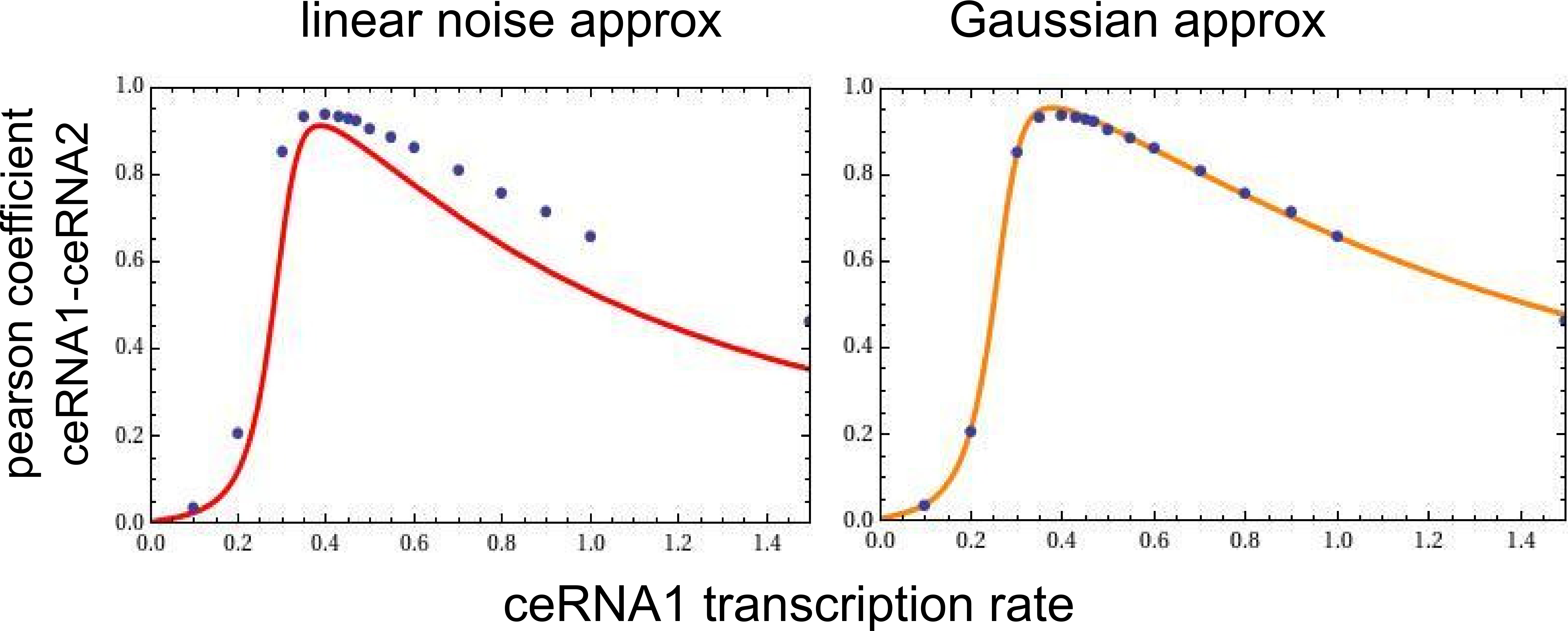}
\caption{{\it Comparison between Linear Noise and Gaussian approximations.} (Left panel) Linear noise approximation, (Right panel) Gaussian approximation. Lines are analytical approximations of the Pearson correlation coefficient. Dots are the results of $10^4$ Gillespie's simulations.}
\label{fig:gaussvslin}
\end{figure}

Thus, the Jacobian and diffusion matrices (${\mathcal A}$ and
${\mathcal B}$ respectively) follow,
{\small
\begin{eqnarray}
{\mathcal A} &=& \left(
\begin{array}{cccc}
-g_{S_1} - \alpha (g_{11} R_1 + g_{12} R_2) & 0 & - \alpha g_{11} S_1 & - \alpha g_{12} S_1 \\
0 & -g_{S_2} - \alpha (g_{21} R_1 + g_{22} R_2) & - \alpha g_{21} S_2 & - \alpha g_{22} S_2 \\
-g_{11} R_1 & -g_{21} R_1 & - g_R - g_{11} S_1 - g_{21} S_2 & 0 \\
-g_{12} R_2 & -g_{22} R_2 & 0 & -g_R - g_{12} S_1 - g_{22} S_2 \\
\end{array}
\right) \nonumber \\
{\mathcal B} &=& \left(
\begin{array}{cccc}
k_{S_1} + g_{S_1} S_1 + \alpha^{2} S_1 A & 0 & \alpha g_{11} R_1 S_1 & \alpha g_{12} R_2 S_1 \\
0 & k_{S_2} + g_{S_2} S_2 + \alpha^{2} S_2 B & \alpha g_{21} R_1 S_2 & \alpha g_{22} R_2 S_2 \\
\alpha g_{11} R_1 S_1 & \alpha g_{21} R_1 S_2 & k_{R_1} + R_1 C & 0 \\
\alpha g_{12} R_2 S_1 & g_{22} R_2 S_2 & 0 & k_{R_2} + R_2 D \\
\end{array}
\right) \; ,
\end{eqnarray}
} with $A = g_{11} R_1 + g_{12} R_2$, $B = g_{21} R_1 + g_{22} R_2$,
$C = g_{R_1} + g_{11} S_1 + g_{21} S_2$ and $D = g_{R_2} + g_{12} S_1
+ g_{22} S_2$.  The covariance matrix elements $c_{rp}$ can be
evaluted accordingly.  In Figure \ref{fig:gaussvslin} we plot the Pearson correlation
coefficient of such system as a function of ceRNA1 transcription
rate. As it is possible to notice, Gaussian approximation performs
better than Linear Noise approximation \cite{TESI_SPAGNOLO}.

\section{Figure parameters (main text)}

\subsection*{Figure 2}

miRNAs transcription rates: $k_{S_1} = 0.05 s^{-1}$ and $k_{S_2} = 0.045 s^{-1}$ ; \\
ceRNA2 transcription rate: $k_{R_2} = 0.155 s^{-1}$ ; \\ 
miRNA degradation rates: $g_{S_1} = g_{S_2} = 0.0002 s^{-1}$ ; \\
ceRNAs degradation rates: $g_{R_1} = g_{R_2} = 0.0004 s^{-1}$ ; \\
ceRNA-miRNA association rates: $g_{11} = g_{12} = g_{21} = g_{22} = 0.0005 s^{-1}$ ; \\
catalyticity parameter: $\alpha = 0.1 $ . \\
ceRNA1 transcription rate is the control parameter and ranges from $0$ to $1.4 s^{s-1}$ .

\subsection*{Figure 3}

miRNA2 transcription rate: $k_{S_2} = 0.03 s^{-1}$ ; \\
ceRNAs transcription rates: $k_{R_1} = 0.355 s^{-1}$ and $k_{R_2} = 0.155 s^{-1}$ ; \\ 
miRNA degradation rates: $g_{S_1} = g_{S_2} = 0.0002 s^{-1}$ ; \\
ceRNAs degradation rates: $g_{R_1} = g_{R_2} = 0.0004 s^{-1}$ ; \\
ceRNA-miRNA association rates: $g_{11} = g_{12} = g_{21} = g_{22} = 0.0005 s^{-1}$ ; \\
catalyticity parameter: $\alpha = 0.1 $ . \\
miRNA1 transcription rate is the control parameter and ranges from $0$ to $0.1 s^{s-1}$ .

\subsection*{Figure 4}

Panel (B):\\
miRNA1-10 transcription rates: $k_{S} = 0.075 + 0.01 \mbox{rand()} s^{-1}$ ; \\
ceRNA2-10 transcription rate: $k_{R} = 0.15 + 0.01 \mbox{rand()} s^{-1}$ ; \\ 
miRNA1-10 degradation rates: $g_{S} = 0.0004 s^{-1}$ ; \\
ceRNAs1-10 degradation rates: $g_{R} = 0.0004 s^{-1}$ ; \\
miRNA-ceRNA association rates: $g = 0.0006$\\
catalyticity parameter: $\alpha = 0.5 $ . \\
ceRNA1 transcription rate is the control parameter and every 35 hours takes the following values: $0.15 s^{-1}$, $0.5 s^{s-1}$, $0.9 s^{s-1}$, $0.15 s^{s-1}$  . \\
\\
Panel (C):\\
miRNA1-9 transcription rate: $k_{S} = 0.02 s^{-1} + 0.01 \mbox{rand()} s^{-1}$ ; \\
ceRNA1-10 transcription rates: $k_{R} = 0.15 + 0.01 \mbox{rand()} s^{-1}$ ; \\
miRNA1-10 degradation rates: $g_{S} = 0.0002 s^{-1}$ ; \\
ceRNA1-10s degradation rates: $g_{R} = 0.0004 s^{-1}$ ; \\
miRNA-ceRNA association rates: $g = 0.0006 s^{-1}$ ; \\
catalyticity parameter: $\alpha = 0.5 $ . \\
miRNA1 transcription rate is the control parameter and every 35 hours takes the following values: $0.02 s^{-1}$, $0.1 s^{s-1}$, $0.8 s^{s-1}$, $0.02 s^{s-1}$  . \\

\subsection*{Figure 5}

miRNAs transcription rates: $k_{S} = 0.2 s^{-1}$ ; \\
ceRNA2 transcription rate: $k_{R_2} = 0.155 s^{-1}$ ; \\ 
miRNA degradation rates: $g_{S_1} = g_{S_2} = 0.0003 s^{-1}$ ; \\
ceRNAs degradation rates: $g_{R_1} = g_{R_2} = 0.0004 s^{-1}$ ; \\
complex association rates: $k_{1}^{+} = k_{2}^{+} = 0.0005 s^{-1}$ ; \\
complex dissociation rates: $k_{1}^{-} = k_{2}^{-} = 0.0003 s^{-1}$ ; \\
complex degradation rates: $\gamma_1 = \gamma_2 = 0.00031 s^{-1}$ ; \\
catalyticity parameter: $\alpha = 0.1 $ . \\
ceRNA1 transcription rate is the control parameter and ranges from $0$ to $1.2 s^{s-1}$ .

\subsection*{Figure 6}

ceRNA1 transcription rates: $k_{R_1} = 0.155 s^{-1}$ ; \\
ceRNA2$_{OFF \rightarrow ON}$ transcription rate jumps from $k_{R_2} = 0$ to $k_{R_2} = 0.345$ ; \\
ceRNA2$_{ON \rightarrow OFF}$ transcription rate jumps from $k_{R_2} = 0.345$ to $k_{R_2} = 0$ ; \\
miRNA degradation rates: $g_{S_1} = g_{S_2} = 0.0002 s^{-1}$ ; \\
ceRNAs degradation rates: $g_{R_1} = g_{R_2} = 0.0004 s^{-1}$ ; \\
ceRNA-miRNA association rates: $g_{11} = g_{12} = g_{21} = g_{22} = 0.0005 s^{-1}$ ; \\
catalyticity parameter: $\alpha = 0.1 $ . \\
miRNA1 transcription rate is the control parameter and ranges from $0$ to $0.5 s^{s-1}$ .
All the other ceRNAs have transcription rates $k_R = 0.1$ and all the other rates equal to ceRNA1 ones.


\section{Response time and experimentally testable trend}

\begin{figure}[!h]
\centering
\includegraphics[scale=0.2]{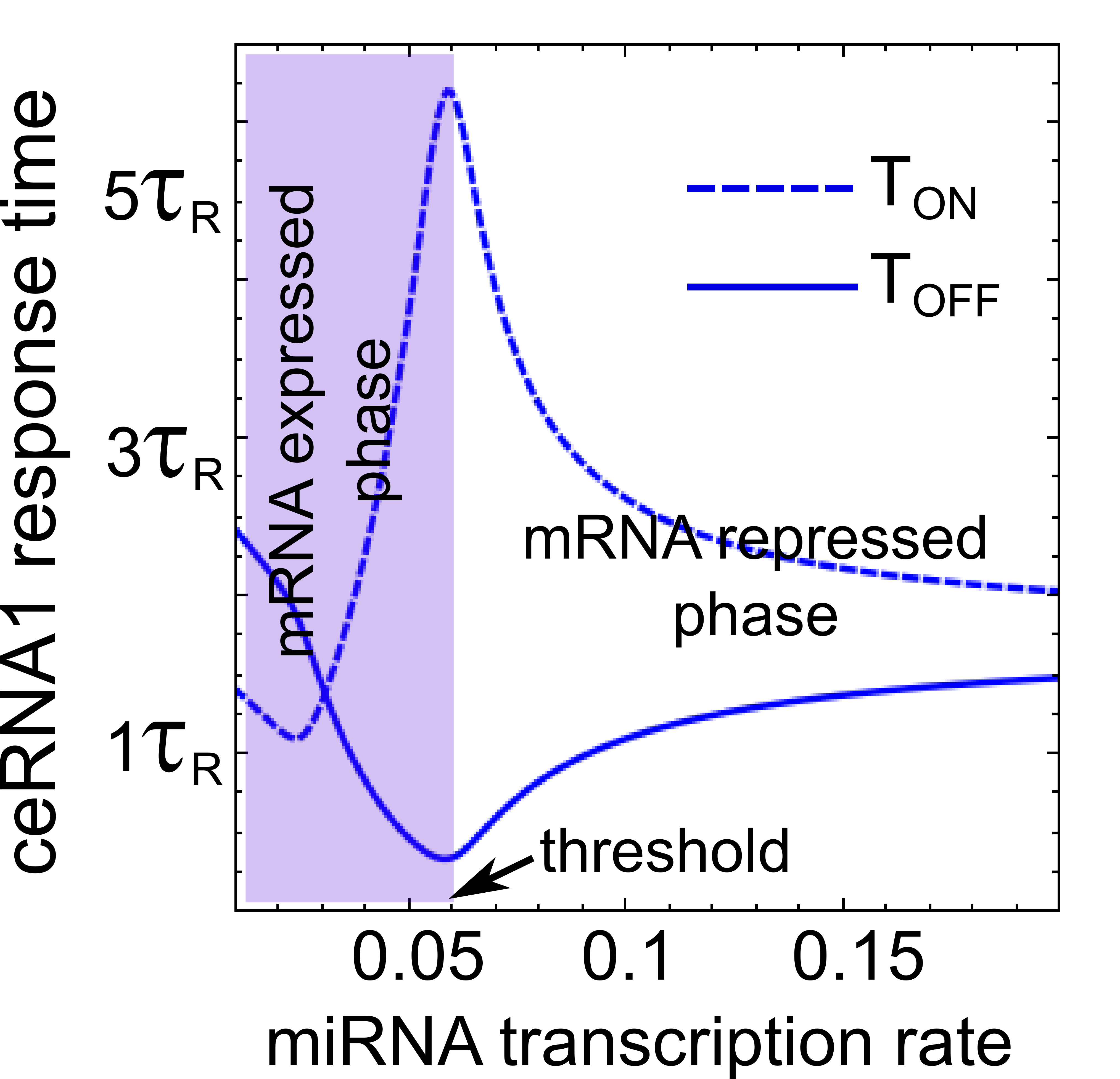}
\caption{{\it Response times and experimentally accessible parameter region.} We show together the $T_{ON}$ and $T_{OFF}$ response times as present in Figure 6 in the main text for the case with 3 ceRNAs. 
The highlighted region corresponds to the experimentally accessible one. Increasing miRNA concentration, switch-off response times show a decreasing trend
while switch-on a U-shaped one. The parameter setting is the same of Figure 6 (main text).}
\label{tempi_SI}
\end{figure}

\bibliography{bibliocerna}

\end{document}